\documentclass[11pt,onecolumn,floatfix,altaffilletter,superscriptaddress,tightenlines,showpacs,showkeys,preprintnumbers,nofootinbib]{revtex4-1}
\pdfoutput=1
\usepackage[colorlinks=true,citecolor=blue,linkcolor=blue,breaklinks=true]{hyperref}
\usepackage{amsmath,amssymb}
\usepackage{epsfig} 
\usepackage{graphicx}        
\usepackage{url}
\usepackage{color}
\usepackage{multirow}
\usepackage{placeins}
\usepackage[dvipsnames]{xcolor}
\usepackage{braket}
\usepackage{float}
\usepackage{slashed}
\usepackage{fdsymbol}
\usepackage{MnSymbol,bbding,pifont}
\hypersetup{
  colorlinks=true,
  linkcolor=Blue,
  filecolor=magenta,   
  urlcolor=blue,
  citecolor=blue,
}

\allowdisplaybreaks

\bibpunct{[}{]}{,}{n}{}{,}

\def\beq{\begin{equation}}
\def\eeq{\end{equation}}
\def\bea{\begin{eqnarray}}
\def\eea{\end{eqnarray}}
\makeatletter
\@addtoreset{equation}{section}

\newcommand{\INFN}{INFN - Sezione di Napoli, Complesso Universitario Monte S. Angelo, I-80126 Napoli, Italy}

\newcommand{\SSM}{Scuola Superiore Meridionale, Università degli studi di Napoli ``Federico II'', Largo San Marcellino 10, 80138 Napoli, Italy}
\newcommand{\NNU}{Department of Physics and Institute of Theoretical Physics,
Nanjing Normal University, Nanjing, 210023, China}
\begin{document}
\title{Cosmic superstrings, metastable strings and ultralight primordial black holes: from NANOGrav to LIGO and beyond}
 \author{Satyabrata Datta}
  \email{amisatyabrata703@gmail.com}
  \affiliation{\NNU}
 \author{Rome Samanta}
  \email{samanta@na.infn.it}
  \affiliation{\SSM}
  \affiliation{\INFN}
\begin{abstract}
   While topologically stable cosmic strings are disfavoured by the recent observation of nHz stochastic gravitational waves (GW) by Pulsar Timing Arrays (PTA), e.g., NANOGrav, cosmic metastable strings and superstrings are not. However, because the gravitational waves from all classes of strings generally span a wide range of frequencies, they contradict LIGO's non-observation of stochastic gravitational waves at the $f\sim $  25 Hz band for a substantial string-parameter space favoured by the PTA data. Suppose ultralight primordial black holes ($M_{\rm BH}<10^9$ g)  existed in the early universe. In this case, they reduce the amplitude of the GWs at higher frequencies by providing an early matter-dominated phase, alleviating the tension between LIGO observation and PTA data. We show that the recent PTA data complemented by future LIGO-Virgo-KAGRA (LVK) runs plus detectors such as LISA and ET would be able to dapple the properties and further search strategies of such ultralight primordial black holes which are otherwise fairly elusive as they evaporate in the early universe by Hawking radiation.
\end{abstract}
\maketitle
\tableofcontents
\section{Introduction}
Ultralight primordial black holes PBHs \cite{Hawking:1971ei,Carr:1974nx,Escriva:2022duf} with initial mass $M_{\rm BH}\lesssim10^8\,g$  are elusive objects as they evaporate in the early universe, therefore, they can not be constrained by astrophysical means like heavier black holes \cite{Carr:2020gox,Sasaki:2018dmp}. Nonetheless, if they existed abundantly and dominated the energy density in the early universe, plenty of direct or indirect observable signatures are achievable in the form of gravitational radiation. For instance, they emit gravitons to constitute a high-frequency gravitational wave background \cite{Anantua:2008am,Dolgov:2011cq,Dong:2015yjs,Ireland:2023avg} and produce density fluctuations to induce GW background testable in the planned GW detectors \cite{Papanikolaou:2020qtd,Domenech:2020ssp,Domenech:2021wkk,Papanikolaou:2022chm,Domenech:2021ztg}. In addition, a PBH domination imprints the spectrum of stochastic gravitational waves background (SGWB) originating from other independent sources. Such imprints on the GWs could be probed with the spectral shape reconstruction techniques \cite{Caprini:2019pxz} planned for the GW detectors. Therefore, a combined study of the GW spectrum associated with the ultralight PBHs--the GW spectrum obtained directly from PBHs plus inquiring into the imprints of a PBH-dominated phase on an independent GW spectrum could be a riveting way to understand the properties of ultralight PBHs. The latter case is the central theme of this article. We study imprints of PBHs on GW background produced by cosmic superstrings \cite{css1,css2,css3,css4,css5,css6,css7,css8,Ellis:2023tsl} and metastable strings \cite{meta1,meta2,meta3,meta4,meta5,meta6,meta7,meta8,Servant:2023tua,Maji:2024pll,Roshan:2024qnv}\footnote{Effect of PBHs on GWs from stable cosmic strings have been recently explored in Refs.\cite{scpbh1,scpbh2}}. We classify the other GW backgrounds obtained directly from PBH evaporation or their dynamics in the early universe as complementary probes that could be envisaged to distinguish the effect of PBH domination on GWs from cosmic strings from any other early matter domination (EMD) scenarios, see, e.g., \cite{Ellis:2023tsl,Antusch:2024ypp}.

The motivation to consider the mentioned classes of strings is twofold. First, because the ultralight PBHs evaporate at early times, before the Big Bang Nucleosynthesis (BBN) \cite{bbn1,bbn2,bbn3,Boccia:2024nly}, they generically affect any propagating GW spectrum during the PBH-domination phase at higher frequencies. Cosmic strings radiate GWs spanning a wide range of frequencies; generally with a strong scale-invariant amplitude at higher frequencies. Therefore, ultralight PBHs leave possibly the cleanest detectable imprints on such GW spectrum. In addition, a wide-spanning GW signal has the following advantage: if a measurement at a certain frequency band fixes the parameter space describing the GW source, signals at other frequency bands get automatically constrained. This redirects us to the second motivation. Cosmic superstrings and metastable strings are speculated to be among the possible sources of nHz SGWB recently discovered by the Pulsar Timing Arrays (PTA) ~\cite{ng1,ng2,ng3,ng4,ng5,EPTA:2023xxk}. Unlike stable cosmic strings \cite{csfit1,csfit2,csfit3,csfit4} described only by the parameter: $G\mu$, where $G$ is the Newton's constant and $\mu\sim v_\Phi^2$ is the string tension with $v_\Phi$ being the vacuum expectation value of the scalar field describing the strings ~\cite{Kibble:1976sj,Hindmarsh:1994re,Jeannerot:2003qv}, cosmic superstrings and metastable cosmic strings fit the data better \cite{Ellis:2023tsl,Figueroa:2023zhu,ng5,metafit1,metafit2,metafit3,metafit4,metafit5,metafit6,metafit7,Pallis:2024mip} because the last two being a two-parameter string model. In addition to $G\mu$, the inter-commutation probability $P$ for superstrings and the parameter $\kappa$ describing the separation between the Grand Unified Theory (GUT) scale and the string scale for metastable strings provide additional degrees of freedom to match the required GW amplitude and spectral slope at nHz frequencies. On the other hand, the high-frequency scale-invariant behaviour of the GW spectrum remains similar for all string classes. Fitting PTA data with GWs from superstrings and metastable strings predicts extremely strong scale-invariant amplitude at higher frequencies if the universe undergoes standard cosmological expansion. In fact, the high-frequency amplitudes are so strong that a significant string parameter space consistent with the PTA data gets ruled out by the LIGO-O3 run which did not observe any SGWB \cite{KAGRA:2021kbb}. This puts superstring and metastable string interpretation to the PTA data perhaps in jeopardy unless an additional GW contribution from supermassive black holes (SMBH) is considered to relax the tension somewhat \cite{ng5}.  A PBH-dominated phase in the early universe, however, can cause a characteristic fall-off of the spectrum at higher frequencies, alleviating the tension with the PTA and LIGO completely. Besides, because the PTA data constrains the string parameters and the PBH mass is bounded from above so that they evaporate before BBN, even though reduced, the predicted overall signal strength could be pretty strong. 

Focusing exclusively on the parameter range of superstrings and metastable strings favoured by the PTA data, and considering non-rotating PBHs with monochromatic initial mass,  we find that PBHs with an initial mass within the range $M_{\rm BH}\in \rm \left[10^6g,10^8 g\right]$ would imprint the GWs at higher frequencies so that they evade the LIGO bound. We broadly classify such viable signals into two classes: whether or not the next LIGO run (LVK-Design) \cite{Aligo1} could probe the PBH-imprinted GW signal obtained from superstrings and metastable strings. We find that for most of the mentioned mass range,  PBHs can produce signals at LVK-D while being consistent with the PTA data. Heavier PBHs with an initial mass around $10^8\,g$ reduce the GW amplitude such that the signals fall short of the sensitivity reach of LVK-D, however strong enough to produce a Signal-to-Noise-Ratio at ET \cite{ET} as large as $\sim 10^4$. 

In the companion theory paper by NANOGrav collaboration \cite{ng5}, the effect of an EMD on GWs from superstrings and metastable strings has been mentioned to evade the LIGO-O3 bound. Therefore, fundamentally, this article does not offer anything new. We rather report if the PBHs provide such EMD, the scenario becomes very distinct and flexile compared to any other EMD, e.g., provided by a scalar field. This is because, first, as mentioned, the PBHs provide additional GW signatures. For the obtained mass range $M_{\rm BH}\in \rm \left[10^6g,10^8 g\right]$, although the GW in the form of gravitons appears at extremely high frequency, GWs, e.g., induced by PBH density fluctuations appear in the detectable range.  It is not trivial to reconstruct such a combined GW signal by any other means.  Second, lately, there has been an extensive effort to understand the effects of ultralight PBHs on beyond the Standard Model (BSM) physics such as particle dark matter (DM) production and baryon asymmetry of the universe (BAU), see, e.g., Refs.\cite{Fujita:2014hha,Lennon:2017tqq,Hooper:2019gtx,Morrison:2018xla} (subsequent references are in Sec.\ref{s4b}). For the $M_{\rm BH}\in \rm \left[10^6g,10^8 g\right]$, such effects could be quite significant, specifically, BSM physics occurring around the Electroweak (EW) scale, even down to the BBN (MeV) scale. In this context, we discuss a recently proposed BSM model \cite{Samanta:2021mdm,Borah:2022iym} accommodating neutrino mass, BAU via leptogenesis, super-heavy DM, and metastable strings (the last feature is new). We show how a PBH-string landscape could be capable of probing untrod parameter regions of BSM models.

\section{Stable strings, superstrings, metastable strings, and non-standard cosmology}\label{s2}
In this section, we briefly discuss/review the foremost aspects of cosmic superstrings and metastable strings required to compute GW spectra in the presence of PBHs. The gravitational wave spectral features obtained from the superstring and metastable string network differ from the standard stable strings mostly at the low frequencies, e.g., in the PTA band. On the other hand, imprints of early matter domination (by PBHs in our case) on the GW spectra at higher frequencies are similar to all string classes. It is therefore useful to highlight first the properties and the technicalities required to compute the GW spectrum from stable cosmic strings. The methodology used can be generalized to the case of superstrings and metastable strings. In addition,  it would be opportune to understand why the GWs from stable cosmic strings do not fit the recent PTA data, unlike the other two string classes.\\

{\bf Stable cosmic strings}: Gravitational waves are radiated from cosmic string loops chopped off from the long strings resulting from the spontaneous breaking of the gauged $U(1)$ \cite{Vilenkin:1981bx,Vachaspati:1984gt}. Long strings are described by a correlation length $L=\sqrt{\mu/\rho_\infty}$, with $\rho_\infty$ is the long string energy density and $\mu $ is the string tension defined as~\cite{Hill:1987qx} 
\bea
\mu=\pi v_\Phi^2~h\left(\lambda,g^\prime\right),\label{tension}
\eea
 $h\left(\lambda,g^\prime\right)\simeq 1$, unless the coupling $\lambda$ and $g^\prime$ are not strongly hierarchical \cite{sh1,sh2}. The time evolution of a radiating loop of initial size $l_i=\alpha t_i$ is given by $l(t)=l_i-\Gamma G\mu(t-t_i)$, where $\Gamma\simeq 50$~\cite{Vilenkin:1981bx,Vachaspati:1984gt}, $\alpha\simeq 0.1$~\cite{Blanco-Pillado:2013qja,Blanco-Pillado:2017oxo}, $G$ is the Newton's constant, and $t_i$ being the initial time of loop production. The total energy loss from a loop can be decomposed into a set of normal-mode oscillations with instantaneous frequencies $f_k=2k/l_k=a(t_0)/a(t)f$, where $k=1,2,3...k_{\rm max}$, $f$ is the present-day frequency at $t_0$, and $a$ is the scale factor. The total GW energy density can be computed by summing all the $k$ modes giving~\cite{Blanco-Pillado:2013qja,Blanco-Pillado:2017oxo} 
\bea
\Omega_{\rm GW}= \sum_{k=1}^{k_{\rm max}}\frac{2k \mathcal{F}_\alpha G\mu^2 \Gamma_k}{f\rho_c}\int_{t_{i}}^{t_0} \left[\frac{a(t)}{a(t_0)}\right]^5 n_\omega\left(t,l_k\right){\rm d}t \,,\label{gwcs1}
\eea
where $\rho_c$ is the critical energy density of the universe, $\mathcal{F}_\alpha\simeq 0.1$ is an efficiency factor~\cite{Blanco-Pillado:2013qja} and $n_\omega\left(t,l_k\right)$ is the loop number density which can be computed from the velocity-dependent-one-scale model as~\cite{Martins:1996jp,Martins:2000cs,Sousa:2013aaa,Auclair:2019wcv}
\bea
n_\omega(t,l_{k})=\frac{A_\beta}{\alpha}\frac{(\alpha+\Gamma G \mu)^{3(1-\beta)}}{\left[l_k(t)+\Gamma G \mu t\right]^{4-3\beta}t^{3\beta}} \,. \label{genn0}
\eea
In Eq.\eqref{genn0}, $\beta=2/3(1+\omega)$ with $\omega$ is the equation of state parameter ({constant}) of the universe, and $A_\beta =5.4$ ($A_\beta = 0.39$) for radiation-dominated (matter-dominated) universe~\cite{Auclair:2019wcv}. {In this section, for illustrative purposes, we shall treat $A_\beta$ as a step function during the change of cosmology plus use Eq.\eqref{genn0} to compute gravitational waves spectra. We shall account for the actual time evolution of $A_\beta$ with VOS equations~\cite{Martins:1996jp,Martins:2000cs,Sousa:2013aaa,Auclair:2019wcv} and appropriate loop number density \cite{ng5} in the next section.} The quantity $\Gamma_k={\Gamma k^{-\delta}}/{\zeta(\delta)}$ quantifies the emitted power in $k$-th mode, with $\delta=4/3$ ($\delta=5/3$) for loops containing small-scale structures such as cusps (kinks) \cite{Damour:2001bk} and makes it evident that  $k=1$ (fundamental) mode is the dominant one contributing to the GWs. In principle, the GW spectrum corresponding to the fundamental mode broadly captures most of the underlying features of a given BSM model producing cosmic strings. However, when it comes to comparing the spectrum with real data, e.g., PTA data, and based on that, one intended to forecast future predictions, it is desirable to sum a large number of modes to obtain accurate results.

The integral in Eq.~\eqref{gwcs1} is subjected to the Heaviside functions $\Theta(t_i-t_{\rm fric})\Theta(t_i-l_{c}/\alpha)$ that set cut-offs on the GW spectrum at high frequencies $f_*$ above which the spectrum falls as $f^{-1}$ for the fundamental mode.  The quantity $t_{\rm fric}$ represents the time until which the string network is damped by friction~\cite{Vilenkin:1991zk}. For $l>l_c$, the GW emission dominates over particle production as shown by recent numerical simulations~\cite{Matsunami:2019fss,Auclair:2019jip}. The critical length can be approximately computed as $l_c\simeq\delta_w (\Gamma G \mu)^{-\gamma}$ where $\delta_w=(\sqrt{\lambda}v_\Phi)^{-1}$ is the width of the string and $\gamma=2$ ($\gamma=1$) for loops containing cusps (kinks). In most cases, the cut-offs owing to friction and kinks are weaker (occur at higher frequencies). We therefore focus on the string loops containing cusps only. 

\begin{figure}[tbp]
\centering 
\includegraphics[scale=0.75]{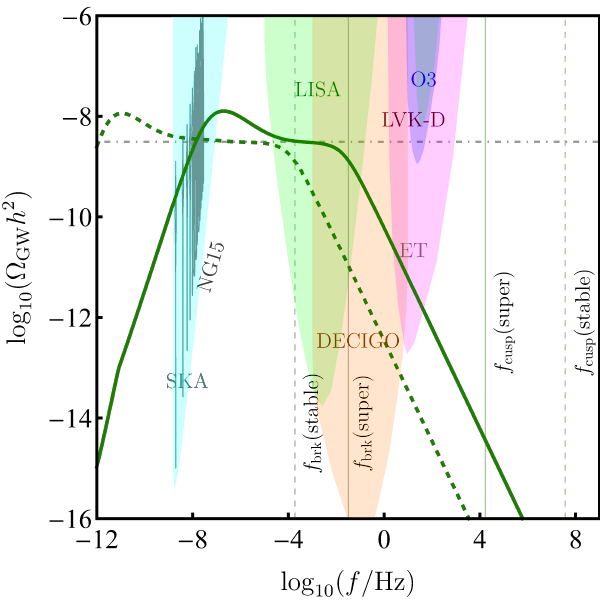}
\includegraphics[scale=0.75]{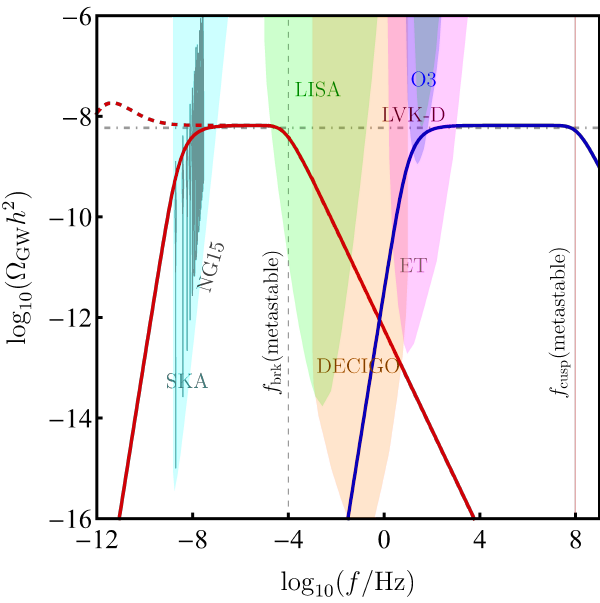}
\includegraphics[scale=1.6]{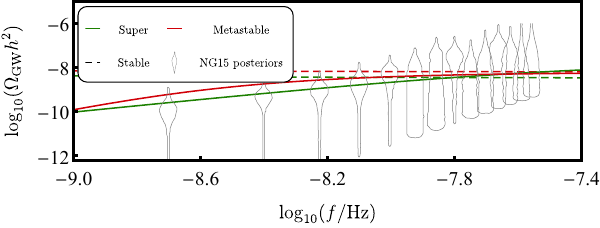}
\caption{\label{fig:1} All GW spectra are produced for $k=1$ mode. Top-left: GW spectra for cosmic super (stable) strings shown with green solid (dashed) curve. Two benchmark string parameters chosen for superstrings are $P=6\times 10^{-3}$ and $G\mu=2\times 10^{-12}$. For the stable strings, we rescale $G\mu$ as $G\mu\rightarrow G\mu/P^2$ so that comparable plateau amplitude is achieved in both cases. The horizontal and vertical dotted black lines represent the analytical expression for the plateau amplitude (multiplied by $P^{-1}$ for superstrings), and the spectral break frequency $f_{\rm brk}$ corresponding to a long-lasting matter domination ending at $T=0.1$ GeV. The green vertical lines correspond to the spectral break frequencies where the spectrum would have fallen due to particle production if there were no matter domination. Top-right: GW spectra for cosmic metastable (stable) strings are shown with the red solid (dashed) curve. In this case, two benchmark string parameters are chosen as $\sqrt{\kappa}=8$ and $G\mu=2\times 10^{-7}$. The dotted horizontal and vertical lines represent the same as in the left panel figure. The red vertical line represents the frequency at which the spectrum would have fallen if there was no matter domination. The GW spectrum represented by the blue curve is for $\sqrt{\kappa}=6$ assuming no matter domination. The density reach of SKA \cite{ska}, LISA \cite{lisa}, DECIGO \cite{decigo} and LVK-D \cite{Aligo1} are shown with the filled coloured curves. Bottom: The benchmark GW spectra in the PTA band with the recent posteriors reported by NANOGrav. }
\end{figure}

  Without any intermediate matter-dominated epoch, Eq.\eqref{gwcs1} can be solved to obtain the GW spectra taking into account two main contributions: i) from the loops created in the radiation domination but emitting in the standard matter domination, ii) from the loops created as well as emitting in the radiation domination. The former corresponds to  a low-frequency peak, while the latter produces a scale-invariant plateau at high frequencies (see Fig.\ref{fig:1}) which for $k=1$ mode is given by~\cite{Blanco-Pillado:2013qja,Blanco-Pillado:2017oxo,Sousa:2020sxs}
\bea
\Omega_{\rm GW}^{\rm plt} = \frac{128\pi\mathcal{F}_\alpha G\mu}{9\zeta(\delta)}\frac{A_R}{\epsilon_R}\Omega_R\left[(1+\epsilon_R)^{3/2}-1\right] \,. \label{flp1}
\eea
 In Eq.~\eqref{flp1}, 
$\epsilon_R=\alpha/\Gamma G\mu \gg 1$, $A_R\equiv A_\beta\simeq 5.4$ and $\Omega_R\sim 9\times 10^{-5}$. Note that the plateau amplitude $\Omega_{\rm GW}^{\rm plt}\propto \sqrt{\mu}\propto v_\Phi$, implying strong amplitude GWs are produced for larger symmetry breaking scale.  In scenarios that feature an early matter-dominated epoch (in this work by PBHs) instead, the plateau breaks at a high frequency $f_{\rm brk}$, beyond which the spectrum falls as $\Omega_{\rm GW}(f>f_{\rm brk})\sim f^{-1}$ for the fundamental mode (green and red curves on the top panel in Fig.\ref{fig:1}). This spectral break frequency $f_{\rm brk}$ can be estimated as  \cite{Cui:2018rwi}  
\bea
f_{\rm brk}\simeq 0.45\,{\rm Hz}\left(\frac{10^{-12}}{G\mu}\right)^{1/2}\left(\frac{T_{\rm brk}}{\rm GeV}\right)\,,\label{brk}
\eea
where $T_{\rm brk}\equiv T_{\rm end} $ is the temperature corresponding to the end of early matter domination. When summed over a large number of modes ($k\rightarrow \infty$), two important changes occur in the GW spectrum. First, the plateau amplitude increases approximately by a factor of  $\zeta(\delta)$, i.e., $\Omega_{\rm GW}^{\rm plt,\infty}\simeq \Omega_{\rm GW}^{\rm plt} \zeta(\delta)$ and the spectrum falls as $f^{1-\delta}$ instead of $f^{-1}$ \cite{Blasi:2020wpy,Schmitz:2024hxw,csfit4,Cheng:2024axj}.

Given an observed spectrum with a plateau followed by a fall-off, one has to compare two parallel scenarios to pinpoint the origin of such a spectrum--whether the EMD or the particle production cut-off causes the spectral shape. Considering the former is equivalent to imposing a constraint on the parameter space as $T_{\rm cusp}/T_{\rm brk}>1$, where we derive $T_{\rm cusp}$  as 
\bea
T_{\rm cusp}\simeq 2.3\times 10^4\,{\rm GeV} \left(\frac{G \mu}{10^{-12}}\right)^{5/4}\,.
\label{c_cusp}
\eea
Similar to Eq.\eqref{brk}, an expression for $f_{\rm cusp}$ can therefore be obtained as 
\bea
f_{\rm cusp}\simeq 1.01\times 10^4\,{\rm Hz} \left(\frac{G \mu}{10^{-12}}\right)^{3/4}\,
\label{f_cusp}
\eea
which we find to be consistent with Eq.(50) of Ref.\cite{Auclair:2019jip}. While analyzing the GW spectrum with PBHs, one has to make sure  that $f_{\rm cusp}>f_{\rm brk}$ is satisfied (see Fig.\ref{fig:1}, top panel).\\

{\bf Cosmic superstrings}: Cosmic superstrings are characterized by a weaker inter-commutation probability $P$ because of their quantum-mechanical nature. The lack of availability of numerical simulations with larger simulation times makes the actual GW spectrum produced by superstrings pretty uncertain. Nonetheless, the standard paradigm to compute the spectrum has been to multiply the loop number density by a factor of $P^{-1}$ \cite{css6,Blanco-Pillado:2017rnf} (see Ref.\cite{css8} for a different scaling relation). This method has also been adopted by the NANOGrav collaboration \cite{ng5}. Therefore, we shall use all the methodologies considered for stable strings but rescaling the loop number density by $P^{-1}$, with $P\in\left[10^{-3},1\right]$ \cite{ng5}. Note that, this is not a rescaling of $G\mu$ that one obtains for hierarchical coupling ($h\left(\lambda,g^\prime\right)\neq 1$) \cite{sh1,sh2}, thereby making the spectral shape, e.g., in the PTA band, different from the stable cosmic strings.  \\

{\bf Metastable cosmic strings}: Metastable cosmic strings decay into string segments connecting monopole-antimonopole 
pairs by quantum tunneling \cite{meta1,meta2}. In the semiclassical approximation, the network is characterized  by the decay rate per string unit length as 
\bea
\Gamma_d = \frac{\mu}{2\pi} \exp(-\pi \kappa),\,\,\sqrt{\kappa } = \frac{m_M}{\mu^{1/2}}\simeq \frac{\Lambda_{\rm GUT}}{v_\Phi},\label{meta}
\eea
where $m_M$  is the monopole mass which is of the order of $\Lambda_{\rm GUT}\simeq 10^{16}$ GeV. From Eq.\eqref{meta}, one can define a time $t_s=\frac{1}{\Gamma_d ^{1/2}}$, before which the cosmic string network behaves similarly to a stable cosmic string network, around which the monopole nucleation starts, and after which, the network behaves as a decaying network with new loops no longer produced. For $t>t_s$ the loop number density in Eq.\eqref{genn0} receives a correction factor \cite{meta6}:
\bea
D(l,t)= e^{-\Gamma_d\left[l(t-t_s)+\frac{1}{2}\Gamma G\mu(t-t_s)^2\right]} \Theta\left(\alpha t_s-\Bar{l}(t_s)\right),
\eea
where the $\Theta$ function with $\Bar{l}(t_s)\simeq l+\Gamma G\mu t$, represents the fact that only the loops produced before $t_s$ contribute to the number density. In principle, unless the $\kappa$ is quite large (which is the case for stable strings), it is sufficient to take the upper limit of the integration in Eq.\eqref{gwcs1} as $t\sim t_{\rm eq}$, because during $t_s<t < t_{\rm eq}$ the loop number density encounters rapid exponential suppression. In this work, we consider GW emission only from loops. Nonetheless, when monopoles do not carry unconfined flux, cosmic string segments formed owing to monopole nucleation can contribute to GWs \cite{meta6}. Compared to the stable cosmic strings, two distinct features of the GW spectrum from metastable strings are i) for smaller values of $\kappa$, the frequency at which the spectrum reaches the plateau gets exponentially right-shifted on the $\Omega_{\rm GW}$ vs. $f$ plane (the blue curve in Fig.\ref{fig:1}). As such, for sufficiently small values of $\kappa$, the spectrum reaches the plateau at a very high frequency $f\gtrsim 10$ Hz, thus avoiding all the existing bound on $G\mu$, barring the one expected from BBN bound on the effective number of degrees of freedom \cite{Peimbert:2016bdg}. ii) The infra-red tail of the GW spectrum behaves as $\Omega_{\rm GW}\propto  f^2$ \cite{meta6,meta7} favored by the PTA data when fitted to a power-law GW spectrum. Markedly, only a few other models exhibit $\Omega_{\rm GW}\propto f^2$ power-law in the infra-red tail, see, e.g., \cite{Babichev:2023pbf,Balaji:2023ehk}. Note that, compared to the stable cosmic strings, in both cases, we have an extra parameter in addition to $G\mu$: the intercommutation probability $P$ for superstrings, and $\kappa$ for metastable strings. This additional degree of freedom (i.e., considering a two-parameter string model) makes it easier to fit the PTA data which are characterized also by two parameters; the amplitude and the spectral index. 

We conclude the discussion by showing the GW spectra in Fig.\ref{fig:1} for superstrings and metastable strings comparing them to the ones from stable strings. We choose a benchmark value of $T_{\rm brk} =0.1$ GeV and consider only the $k=1$ mode, leaving a more realistic analysis by summing a large number of modes for the next section. In Fig.\ref{fig:1} (left) we show the GW spectrum for cosmic super (stable) strings with the green solid (dashed) curve. For the superstrings, we choose the benchmark values $P=6\times 10^{-3}$ and $G\mu=2\times 10^{-12}$ while for the stable cosmic strings, we rescale $G\mu$ as $G\mu\rightarrow G\mu/P^2$ such that comparable plateau amplitude can be obtained in both cases. Note that, for superstrings, one has the required blue-tilted GW spectrum in the PTA band (A blue-tilted distribution of fourteen NANOGrav posteriors is shown in gray; in this work, we use only NANOGrav data which are consistent with all other PTAs (IPTA) \cite{ipta}). On the other hand, even though the required GW amplitude can be obtained for the stable strings, the spectrum is nearly scale-invariant in the  PTA band and thereby does not fit the data. A similar situation occurs in the case of metastable strings as shown in the top right panel with the red curves\footnote{With the blue curve we show the GW spectra for $\sqrt{\kappa}=6$ evading all the bounds with the dominant contribution at very high-frequencies. Although this spectrum is not so relevant for the present study, we find it interesting to mention it as it is generally not present in the case of stable (and super) strings.}. In both plots, the black dotted lines represent the analytical expression for the plateau amplitude and the spectral break frequency given in Eq.\eqref{flp1} and \eqref{brk}, respectively.  The green and the red vertical lines represent the spectral break frequency $f_{\rm cusp}$ at which the spectrum would have fallen if there was no matter domination. In the bottom panel, we zoomed in on the GW spectra within the PTA band to show the consistency of the blue-tiled GW spectrum obtained from cosmic superstrings and metastable strings with the data.

An important aspect of this study is considering a falling GW spectrum at higher frequencies to avoid the LIGO bound. Although we zero in on early matter domination for this purpose, as discussed earlier, the spectrum may also fall because of particle production. With a straightforward generalization of Eq.(11) of Ref.\cite{Matsunami:2019fss} with $\delta_w=(\sqrt{\lambda} v_\Phi)^{-1}$, it can be shown that for making the spectrum fall before the LIGO frequency band while being consistent with the PTA data, an extremely small value of $\lambda$ is required. For the standard case with $\lambda\sim 1$ considered in the numerical simulations, the spectrum always falls at very high frequency owing to particle production. This makes the case of an early matter-dominated scenario pragmatic while interpreting the PTA data as cosmic metastable (super) strings radiated GWs evading the LIGO-O3 bound.

\section{Imprints of ultralight PBHs on GWs from cosmic (super/metastable) strings}\label{s3}

We focus on non-rotating PBHs with monochromatic initial mass $M_{\rm BH}$. The energy density of the PBHs ($\rho_{\rm BH}$) and radiation ($\rho_R$) evolve according to the following Friedmann equations \cite{Giudice:2000ex,Masina:2020xhk,csfit4}:
\bea
\frac{d\rho_{R}}{dz}+\frac{4}{z}\rho_R&=&0,\label{den1}\\
\frac{d\rho_{\rm BH}}{dz}+\frac{3}{z}\frac{H}{\tilde{H}}\rho_{\rm BH}&-&\frac{\dot{M}_{\rm BH}}{M_{\rm BH}}\frac{1}{z\tilde{H}}\rho_{\rm BH}=0,\label{den2}
\eea
where $H$ is the Hubble parameter ($H_0\sim 1.44\times 10^{-42}$ GeV),  $z=T_{\rm Bf}/T$, with $T_{\rm Bf}$ being the PBH formation temperature and the time derivative $\dot{M}_{\rm BH}$ is given by
\bea
\dot{M}_{\rm BH}=\frac{\mathcal{G}g_{*B}(T_{\rm BH})}{30720\pi}\frac{M_{\rm Pl}^4}{M_{\rm BH}^2},\label{deri}
\eea
where $M_{\rm Pl}=1.22\times 10^{19}$ GeV, $\mathcal{G}\simeq 3.8$ is a grey body factor and $g_{*B}(T_{\rm BH})\simeq 100$ being the number of relativistic degrees of freedom below the PBH temperature $T_{\rm BH}$ with Standard Model (SM) containing left-handed neutrinos \cite{MacGibbon:1991tj}. The quantity $\tilde{H}$ and the scale factor $a$ evolve as
\bea
\tilde{H}=\left(H+\mathcal{K}\right),~
\frac{da}{dz}=\left(1-\frac{\mathcal{K}}{\tilde{H}}\right)\frac{a}{z},
\label{temvar}
\eea
where $\mathcal{K}=\frac{\dot{M}_{\rm BH}}{M_{\rm BH}}\frac{\rho_{\rm BH}}{4\rho_{R}}$. To derive Eq.\eqref{den1}-Eq.\eqref{temvar}, we assume the entropy ($g_{*s}$) and the energy ($g_{*\rho}$) degrees of freedom are equal and constant. For a given value of $\beta\equiv \frac{\rho_{\rm BH}(T_{\rm Bf})}{\rho_{\rm R}(T_{\rm Bf})}$ and $M_{\rm BH}$,  the above equations can be solved to determine the duration of PBH domination and the evaporation temperature. In the case where the PBH dominates the energy density, an  analytical expression for the evaporation temperature is obtained as 
\bea
T_{\rm ev}=\left(\frac{5 M_{\rm Pl}^2}{\pi^3g_{*}(T_{\rm ev})\tau^2}\right)^{1/4},
\eea
where $\tau$ is the PBH lifetime which can be computed by integrating Eq.\eqref{deri}. This allows also to express $T_{\rm ev}$ in terms of $M_{\rm BH}$ as
\bea
T_{\rm ev}=0.1\left(\frac{M_{\rm BH}}{4.1\times 10^7\,g}\right)^{-3/2}\, {\rm GeV}.\label{pbheva}
\eea
Eq.\eqref{pbheva}  shows that with $T_{\rm ev}\equiv T_{\rm brk}$, all the plots in Fig.\ref{fig:1} can be reproduced for PBHs with initial mass $M_{\rm BH}\simeq 4.1\times 10^7\, g$.

It is crucial to note the following: i) An eternal falling of the GW spectrum in the range $f_{\rm brk}\lesssim f\lesssim f_{\rm LIGO}$, (see Fig.\ref{fig:1}), requires longer duration of PBH domination and therefore large $\beta$--we consider this to constrain PBH mass with PTA data and the LIGO bound plus projection. However, in scenarios where $\beta$ can be constrained (can be bounded from above, resulting in a short duration of PBH domination) by GWs from PBH density fluctuations \cite{Domenech:2020ssp}, the falling of the GW spectrum might stop before the LIGO frequency band and another subsequent plateau with suppressed amplitude (originating out of the contribution from the loops in the first radiation epoch before the PBH start to dominate) might show up. Therefore one needs to modify our results accordingly, see, e.g., sec.\ref{s4}. ii) The PBHs should evaporate before the BBN at a temperature $T\gtrsim 10$ MeV. Therefore, from Eq.\eqref{pbheva}, $M_{\rm BH}$ is generically constrained as $M_{\rm BH}\lesssim 2\times 10^8\,g$, which we use as the absolute upper bound on $M_{\rm BH}$. 

We now proceed to the detailed discussion of the impact of PBH-induced matter domination on GWs from cosmic superstrings and metastable strings. First, from Eq.\eqref{brk} and Eq.\eqref{pbheva}, we obtain the spectral break frequency as
\bea
f_{\rm brk}^{\rm PBH}\simeq 0.045\,{\rm Hz}\left(\frac{10^{-12}}{G\mu}\right)^{1/2}\left(\frac{M_{\rm BH}}{4.1\times 10^7\,g}\right)^{-3/2}\,.\label{brkpbh}
\eea
{
Note that the expression for $ f_{\rm brk}^{\rm PBH} $ is strictly valid under the assumption of an instantaneous growth of $ A_\beta $ (from 0.39 to 5.4) as the string network transitions from the PBH-dominated epoch to the radiation-dominated era and the loop number density given in Eq.\eqref{genn0} which is a special case of the generic expression of $n_\omega(t,l_{k})$ given in Eq.(43) of Ref.\cite{ng5} under the assumption that the equation of state does not change in time. In reality, either of these is not true during the transition period and must be appropriately considered in the numerical computation.

The correct loop number density and the gradual growth of  $ A_\beta $ (this parameter is defined as $C(t)$ in Ref.\cite{ng5} in Eq.(45)) approaching its instantaneous value can be tracked accurately using Eq.\eqref{den1},  Eq.\eqref{den2}, and Eq.\eqref{temvar} (note that this equation tracks the time evolution of the scale factor to be fed in Eq.(43) of Ref.\cite{ng5}) and the velocity-dependent one-scale (VOS) equations of motion~\cite{Martins:1996jp,Martins:2000cs,Sousa:2013aaa,Auclair:2019wcv}:

\bea
\frac{dL}{dt}=HL(1+\tilde{v})^2+\frac{1}{2}\tilde{c}\tilde{v},\label{vos1}\\
\frac{d\tilde{ v}}{dt}=1-\tilde{v}^2\left(\frac{k(\tilde{v})}{L}-2H\tilde{v}\right),\label{vos2}
\eea
where $\tilde{v}$ is the root-mean-squared (RMS) velocity of the long strings, $c=0.23\pm 0.04$ and the function $k(\tilde{v})$ which phenomenologically accounts for the effects of small structures on the long strings is given by
\bea
k(\tilde{v})=\frac{2\sqrt{2}}{\pi}(1-\tilde{v}^2)(1+2\sqrt{2}\tilde{v}^3)\frac{1-8\tilde{v}^6}{1+8\tilde{v}^6}.
\eea
Parametrising $L=\xi (t)t$ one defines $A_\beta$ as $A_\beta=\frac{\tilde{c}\tilde{v}}{\sqrt{2 }\xi^3}$, where $\xi(t)=\alpha(t)/\alpha_L$ with $\alpha_L\simeq 0.37$. So far we have assumed that Eq.\eqref{vos1} and Eq.\eqref{vos2} reach their attractor solutions ($\tilde{v}$ and $\xi$ are constant) instantly during the change of cosmology. In our case, precise evolution of $A_\beta(z)$ can be obtained upon recasting Eq.\eqref{vos1} and Eq.\eqref{vos2} as
\bea
\frac{dt}{dz}&=&\frac{1}{\tilde{H}z},\\
\frac{d\xi}{dz}&=&\frac{1}{t\tilde{H}z}\left(H t\xi(1+\tilde{v})^2+\frac{1}{2}\tilde{c}\tilde{v}-\xi\right),\\
\frac{d\tilde{v}}{dz}&=&\frac{1-\tilde{v}^2}{\tilde{H}z}\left(\frac{k(\tilde{v})}{\xi t}-2H\tilde{v}\right)
\eea
and solving them together with Eq.\eqref{den1} and Eq.\eqref{den2}.
\begin{figure}
\centering 
\includegraphics[scale=1]{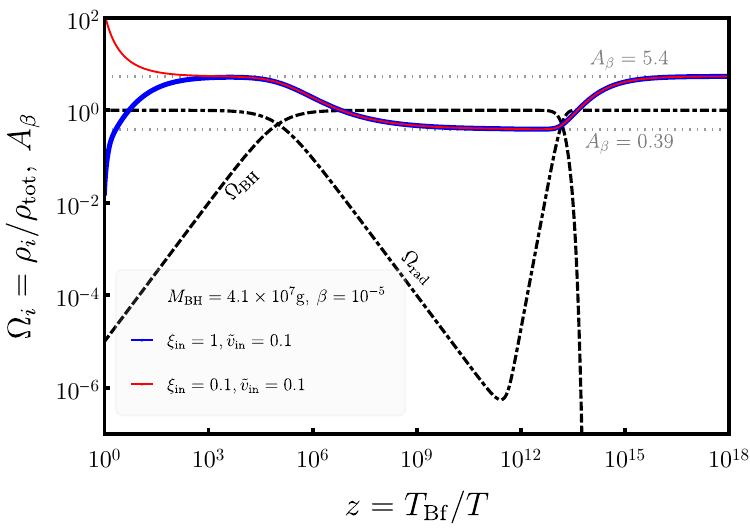}
\caption{Evolution of normalized energy densities and $A_\beta$ with $z$ in presence of ultralight PBH with initial mass $M_{\rm BH}=4.1\times 10^7$ g. The blue and the red lines representing $A_\beta$ correspond to two different initial conditions.}\label{vosfig}
\end{figure}
In Fig.~\ref{vosfig}, we present the evolution of $ A_\beta $ as a function of $ z $ for two distinct initial conditions: $ \tilde{v}_{\rm in} = 0.1 $ and $ \xi_{\rm in} = 0.1, 1 $, represented by the red and blue curves, respectively. The figure clearly illustrates that, in both scenarios, $ A_\beta $ eventually converges to the attractor solution. However, the transitions are gradual in both cases: the shift from the initial radiation epoch to the PBH-dominated epoch occurs around $ z \sim 10^5 $, followed by a transition from the PBH-dominated epoch to the latest radiation era near $ z \sim 10^{14} $. The later transition is important in our discussion. 

After appropriately taking into account the loop number density (Eq.(43) of Ref.\cite{ng5}) and variation of $A_\beta$ presented in Fig.\ref{vosfig}, we find that an approximate analytical relation for the spectral break frequency can be written as
\bea
f_{\rm brk}^{\rm PBH}\simeq 0.008\,{\rm Hz}\left(\frac{10^{-12}}{G\mu}\right)^{1/2}\left(\frac{M_{\rm BH}}{4.1\times 10^7\,g}\right)^{-3/2}\,\label{brkpbhnew}
\eea
accounting for a percent level deviation from the near-scale-invariant plateau. For further computations, we will use the spectral break frequency defined in Eq.~\eqref{brkpbhnew}, which is smaller by order of magnitude compared to the instantaneous case shown in Eq.~\eqref{brkpbh}. 

 }

When a large number of modes are summed, GW energy density beyond $f_{\rm brk}^{\rm PBH}$ can be written as 
\bea
\Omega_{\rm GW}(f)\simeq \Omega_{\rm GW}^{\rm plt,\infty} \left(\frac{f}{f_{\rm brk}^{\rm PBH}}\right)^{-1/3}\Theta\left[ f-f_{\rm brk}^{\rm PBH}\right],\label{gwpbh}
\eea
where 
\bea
 \Omega_{\rm GW}^{\rm plt,\infty} \simeq 3.1\times 10^{-8}\,\Upsilon^{-1}\left(\frac{G\mu}{10^{-7}}\right)^{1/2},\,\,\begin{cases}
     \Upsilon=P\,\,{\rm for \,\, superstrings}\\ \Upsilon=1\,\,{\rm for \,\, metastable\, strings.}
 \end{cases}\label{lcons}
\eea
Using Eq.\eqref{brkpbh}, Eq.\eqref{gwpbh} can be re-expressed as 
\bea
\Omega_{\rm GW}(f)\simeq \Omega_{\rm GW}^{\rm plt,\infty}  \Omega_{\rm GW}^{\rm mod,PBH},
\eea
where $\Omega_{\rm GW}^{\rm mod,PBH}$, given by
{
\bea
\Omega_{\rm GW}^{\rm mod,PBH}\simeq 0.068\,\left(\frac{10^{-12}}{G\mu}\right)^{1/6}\left(\frac{M_{\rm BH}}{4.1\times 10^7\,g}\right)^{-1/2}\left(\frac{f}{25\,\rm Hz}\right)^{-1/3}
\eea
}
 captures the modulation in the GW spectra because of PBH domination. The absence of PBHs corresponds to $\Omega_{\rm GW}^{\rm mod, PBH}=1$. In which case, Eq.\eqref{lcons} and the non-observation of SGWB by LIGO-O3 run; i.e., $\Omega_{\rm GW}^{\rm any\,model}\lesssim 1.7\times 10^{-8}$ at $f_{\rm LIGO}=25$ Hz, implies the following: 
\bea
{\rm without\, the\, PBH,\, LIGO-O3\,excludes\,\,
}\begin{cases}
    P\lesssim 1.8\,\left(\frac{G\mu}{10^{-7}}\right)^{1/2}\,\,{\rm for \,\, superstrings,}\\
    G\mu\gtrsim 3\times10^{-8}\,\,{\rm for \,\, metastable\,strings.}
\end{cases}\label{lcons2}
\eea

\begin{figure}
    \centering
    \includegraphics[scale=0.75]{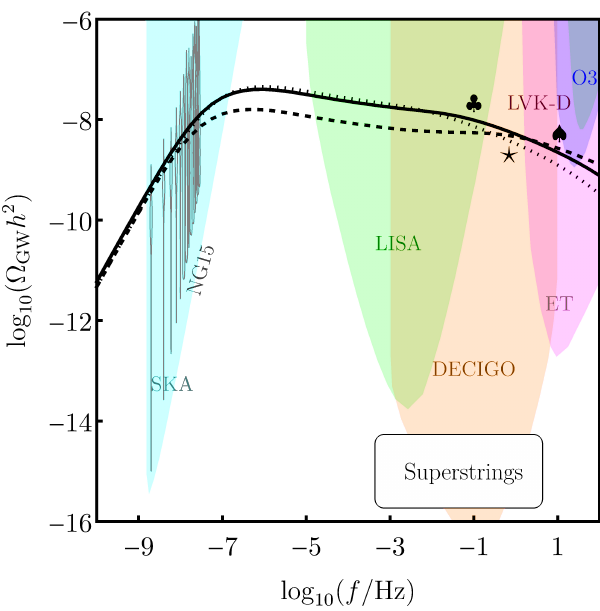}\includegraphics[scale=0.75]{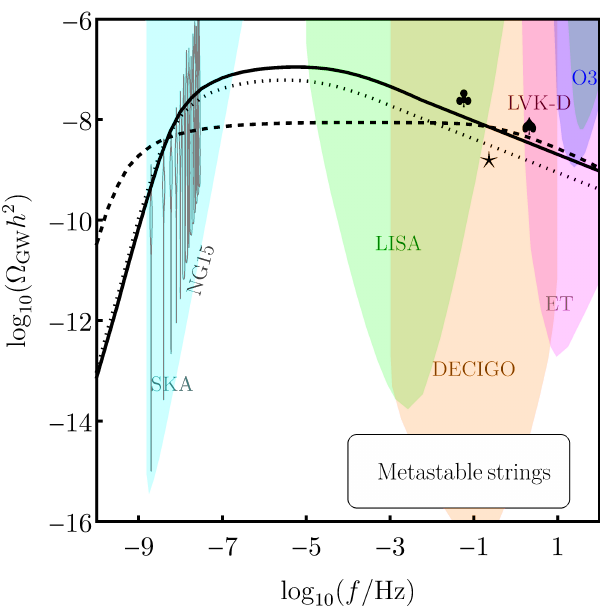}
    \caption{When PBHs dominate in the early universe: Left (right): Gravitational wave spectra for three benchmark points [see Table \ref{t1}, \ref{t2}] from each category discussed in the text are shown for superstrings (metastable strings). We consider GWs radiated from loops only for metastable strings, discarding contribution from string segments. Each benchmark GW spectrum is consistent with PTA data and evades LIGO-O3 bound on stochastic gravitational waves. {\color{black}To produce these GW spectra and all the other spectra in subsequent figures we sum $10^3$ modes and the extrapolate it to higher frequencies keeping $f^{-1/3}$ falling.}}
    \label{fig:fig2}
\end{figure}
Though Eq.\eqref{lcons2} is the most interesting region to explore with non-standard cosmological evolution, e.g., with PBHs, for a comprehensive discussion, we present our numerical results by dividing parameter space into three different categories (all cases are consistent with PTA data):\\

Category A: $\Omega_{\rm GW}^{\rm plt,\infty} \gtrsim \Omega_{\rm GW}^{\rm LIGO-O3} $ and $\Omega_{\rm GW}^{\rm LIGO-O3} \gtrsim \Omega_{\rm GW}^{\rm plt,\infty}  \Omega_{\rm GW}^{\rm mod,PBH}(f=25\,{\rm Hz})\gtrsim \Omega_{\rm GW}^{\rm LVK-D}$.\\

Category B: $\Omega_{\rm GW}^{\rm LIGO-O3}\gtrsim\Omega_{\rm GW}^{\rm plt,\infty} \gtrsim \Omega_{\rm GW}^{\rm LVK-D}$ and $\Omega_{\rm GW}^{\rm LIGO-O3}\gtrsim\Omega_{\rm GW}^{\rm plt,\infty}\Omega_{\rm GW}^{\rm mod,PBH}(f=25\,{\rm Hz}) \gtrsim \Omega_{\rm GW}^{\rm LVK-D}$.

Category C: $\Omega_{\rm GW}^{\rm LVK-D}\gtrsim\Omega_{\rm GW}^{\rm plt,\infty}\Omega_{\rm GW}^{\rm mod,PBH}(f=25\,{\rm Hz})$ for all $\Omega_{\rm GW}^{\rm plt,\infty}$.\\

Here $\Omega_{\rm GW}^{\rm LVK-D}=2\times 10^{-9}$ at $f=25$ Hz is the maximum projected sensitivity (minimum testable value of $\Omega_{\rm GW}$ ) reach of the LVK-D. Note that Category A corresponds to recovering the excluded parameter space in Eq.\eqref{lcons2}, for which one must require a non-standard cosmological evolution, e.g., by PBH. For Category B, PBHs are not essential because even the scale-invariant amplitude does not contradict the LIGO-O3 bound. A PBH domination however can be tested for $f_{\rm brk}^{\rm PBH}\lesssim25\,{\rm Hz}$ and $\Omega_{\rm GW}^{\rm LIGO-O3}\gtrsim\Omega_{\rm GW}^{\rm plt,\infty}\Omega_{\rm GW}^{\rm mod,PBH}(f=25\,{\rm Hz}) \gtrsim \Omega_{\rm GW}^{\rm LVK-D}$. For Category C, though $\Omega_{\rm GW}$ is consistent with the PTA data, the LVK-D cannot test the spectrum because the amplitude is smaller than its sensitivity reach. However, detectors like the Einstein Telescope (ET) can potentially test such signals. We quantify this class of signals by computing Signal-to-Noise-Ratio at ET. In Fig.\ref{fig:fig2}, we show the benchmark GW spectrum (summing over larger modes) from each category belonging to both string classes. The benchmark parameters are tabulated in Table \ref{t1} (Superstrings) and Table \ref{t2} (Metastable strings). In what follows, we present detailed numerical results including a more rigorous parameter space scan to find an approximate mass range of PBHs that can produce such signals. 
\begin{table}[t!]
    \centering 
    \begin{minipage}{.5\linewidth}
     \caption{Benchmarks: Superstrings }\label{t1}
    \begin{tabular}{c|c|c|c}
         Benchmark cases\,& $~~G\mu~~$ & $P$ & $~~~M_{\rm BH}~[{\rm g}]~~~$\\ \hline
         BP1 (Cat-A) & $ 10^{-11.6}$ & $~~ 10^{-2.3}~~$ & $ 10^{7.5}$\\
         BP2 (Cat-B) & $ 10^{-11.4}$ & $~~ 10^{-1.8}~~$ & $ 10^{6.5}$\\
         BP3 (Cat-C) & $ 10^{-11.7}$ & $~~ 10^{-2.4}~~$ & $ 10^{8}$\\
        
    \end{tabular}
    \end{minipage}%
    \begin{minipage}{.5\linewidth}
    \caption{Benchmarks: Metastable strings }\label{t2}
    \begin{tabular}{c|c|c|c}
         Benchmark cases\,& $~~G\mu~~$ & $\sqrt{\kappa}$ & $~~~M_{\rm BH}~[{\rm g}]~~~$\\ \hline
         BP1 (Cat-A) & $ 10^{-5}$ & $~~ 7.85~~$ & $ 10^{7}$\\
         BP2 (Cat-B) & $ 10^{-7.4}$ & $~~ 8.2~~$ & $ 10^{5.5}$\\
         BP3 (Cat-C) & $ 10^{-5.5}$ & $~~ 7.9~~$ & $ 10^{7.5}$\\
        
    \end{tabular}
    \end{minipage}
    \label{tab:BP}
\end{table}
\subsection{Numerical results: PBH mass ranges for PTA GW signals evading LIGO-O3 bound }
\begin{figure}
    \centering
    \includegraphics[scale=0.58]{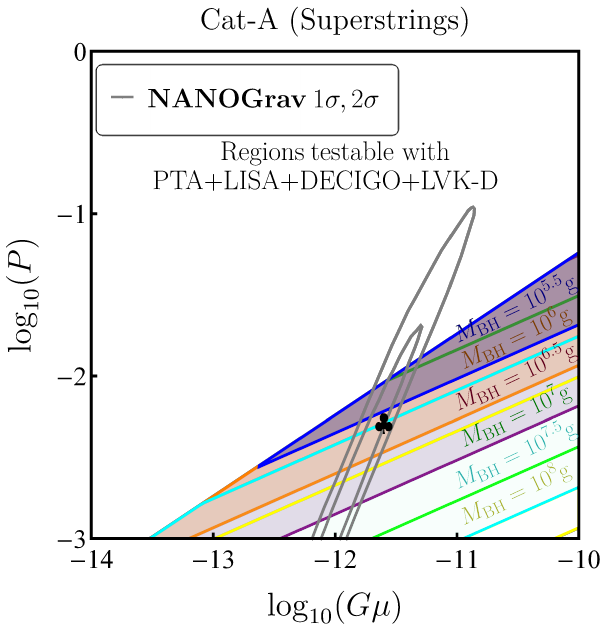}\includegraphics[scale=0.53]{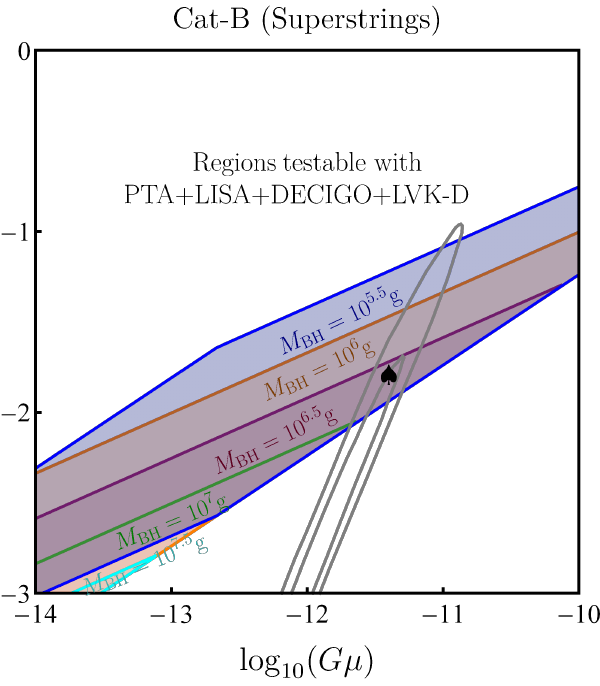}\includegraphics[scale=0.53]{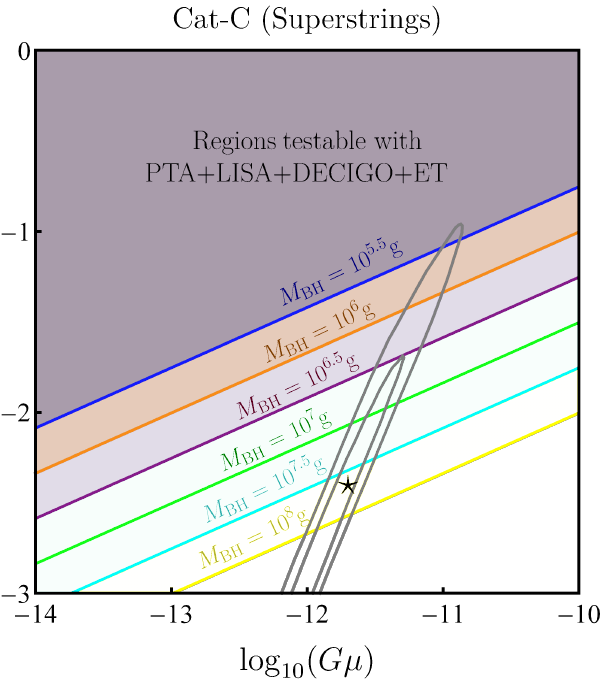}
    \includegraphics[scale=1.1]{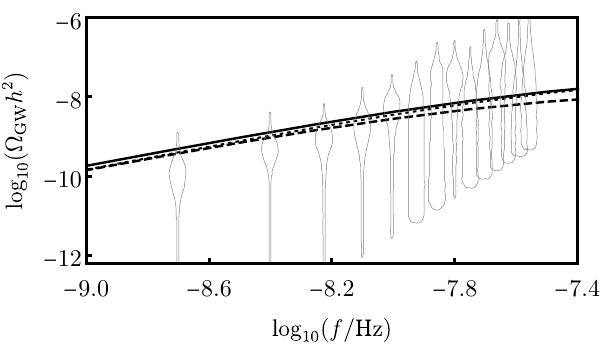}\includegraphics[scale=0.95]{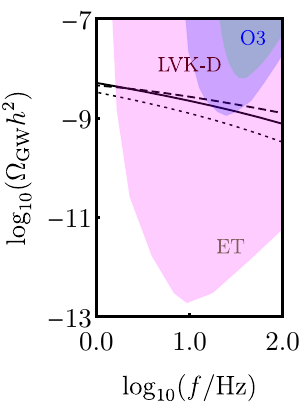}
    \caption{All plots are for superstrings. Top panel, left: The coloured regions marked with different PBH mass $M_{\rm BH}$ on the $G\mu-P$ plane, correspond to the GW spectra with plateau amplitude $\Omega_{\rm GW}^{\rm plt,\infty} \gtrsim \Omega_{\rm GW}^{\rm LIGO-O3}\equiv 1.7\times 10^{-8} $ and testable with LVK-D at 25 Hz. In the absence of PBHs, the regions would be excluded by LIGO-O3 because in that case, the GW spectra do not fall at $f\ll25$ Hz, and the scale-invariant amplitudes contradict the LIGO bound.  The regions inside the gray contours are consistent with the NANOGrav 15yrs data. Middle: The colored regions correspond to the GW spectra $\Omega_{\rm GW}^{\rm LIGO-O3}\gtrsim\Omega_{\rm GW}^{\rm plt,\infty} \gtrsim \Omega_{\rm GW}^{\rm LVK-D}\equiv 2\times 10^{-9}$. Even in the absence of PBHs, these regions do not contradict the LIGO-O3 bound. But in that case, LVK-D would probe a scale-invariant spectrum. Right: These regions correspond to GWs with smaller amplitudes at the LIGO frequency band, so LVK-D can not test them, but detectors such as ET can. For the chosen benchmark point from each category (Table. \ref{t1}) the GW spectrum is shown in Fig.\ref{fig:fig2} (left). Bottom panel: Zoomed in GW spectra in the PTA band against the NANOGrav posteriors (the violins) and in the LIGO frequency band.}
    \label{fig:fig3}
\end{figure}

\begin{figure}
    \centering
    \includegraphics[scale=0.58]{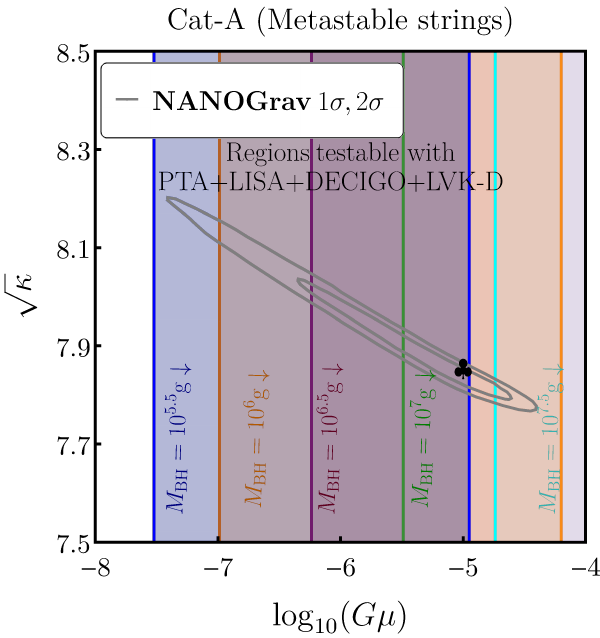}\includegraphics[scale=0.53]{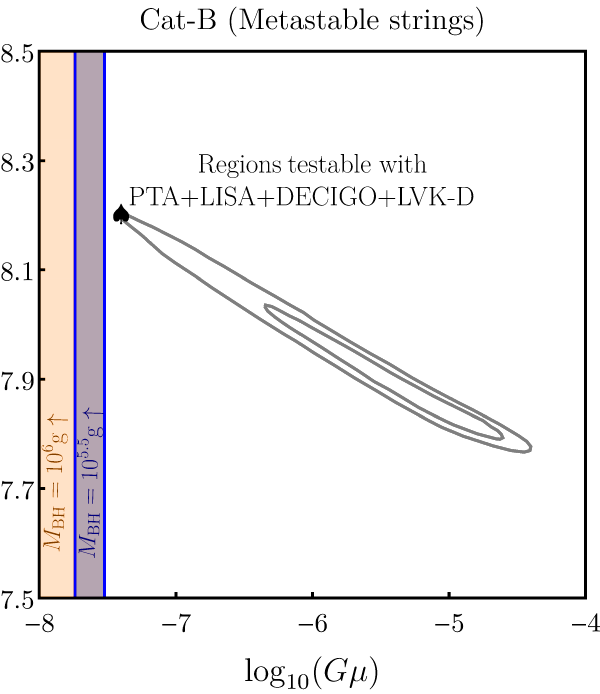}\includegraphics[scale=0.53]{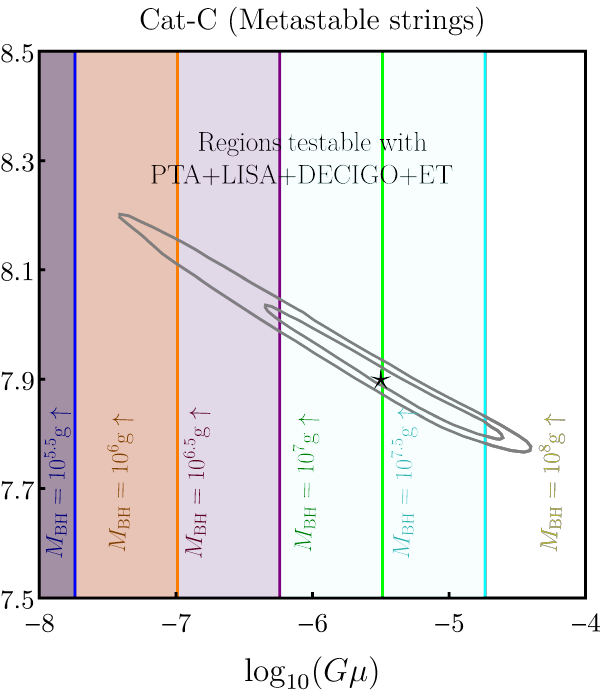}\\
     \includegraphics[scale=1.1]{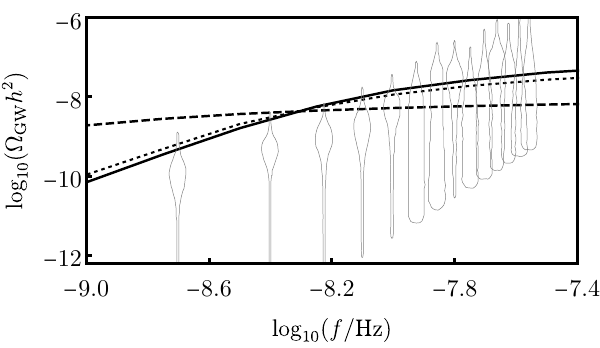}\includegraphics[scale=0.95]{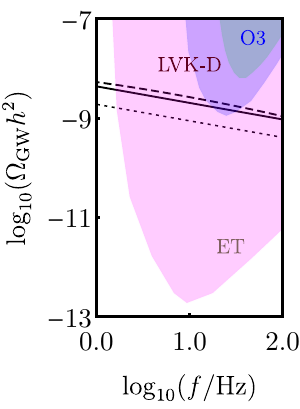}
   \caption{All plots are for metastable strings. Top panel, left: The coloured regions marked with different PBH mass $M_{\rm BH}$ on the $G\mu-\kappa$ plane, correspond to the GW spectra with plateau amplitude $\Omega_{\rm GW}^{\rm plt,\infty} \gtrsim \Omega_{\rm GW}^{\rm LIGO-O3}\equiv 1.7\times 10^{-8} $ and testable with LVK-D at 25 Hz. In the absence of PBHs, the regions would be excluded by LIGO-O3 because in that case, the GW spectra do not fall at $f\ll25$ Hz, and the scale-invariant amplitudes contradict the LIGO bound.  The regions inside the gray contours are consistent with the NANOGrav 15yrs data. Middle: The colored regions correspond to the GW spectra $\Omega_{\rm GW}^{\rm LIGO-O3}\gtrsim\Omega_{\rm GW}^{\rm plt,\infty} \gtrsim \Omega_{\rm GW}^{\rm LVK-D}\equiv 2\times 10^{-9}$. Even in the absence of PBHs, these regions do not contradict the LIGO-O3 bound.  Right: These regions correspond to GWs with smaller amplitudes at the LIGO frequency band, so LVK-D can not test them, but detectors such as ET can. For the chosen benchmark point from each category (Table \ref{t2}) the GW spectrum is shown in Fig.\ref{fig:fig2} (right). Bottom panel: Zoomed in GW spectra in the PTA band against the NANOGrav posteriors (the violins) and in the LIGO frequency band.}
    \label{fig:fig4}
\end{figure}

On the top panel in Fig.\ref{fig:fig3}, we show the allowed regions for fixed PBH mass that would produce GW signals in each category belonging to cosmic superstrings. The gray contours in each plot are the NANOGrav $1\sigma$ and $2\sigma$ posteriors of $G\mu$ and $P$ \cite{ng5} (one may also look at the Appendix \ref{a1} of this article). The benchmark points tabulated in Table \ref{t1} corresponding to the GW spectra in Fig.\ref{fig:fig2} (left) are denoted with the symbols $\clubsuit$  BP1 (Cat-A), $\spadesuit$  BP2 (Cat-B), and \scalebox{2}{$\thinstar$}  BP3 (Cat-C). A similar description holds for metastable strings on the $G\mu$ and $\sqrt{\kappa}$ plane (Fig.\ref{fig:fig4}, top) with the benchmarks tabulated in Table \ref{t2} and corresponding GW spectra shown in Fig.\ref{fig:fig2} (right). In the bottom panel of both figures, the GW spectra presented in Fig.\ref{fig:fig2} are zoomed in to the NANOGrav and the LIGO frequency band.  Note that for all categories in both figures, only a constrained range of PBH mass $M_{\rm BH}\sim \left[10^{6}\,g,\,10^8\,g\right]$ would produce a viable GW signal consistent with the PTA data. For category C, slightly heavier PBHs are allowed; to make the spectrum fall below $\Omega_{\rm GW}^{\rm LVK-D}$ with $f_{\rm brk}$  at lower frequencies. On the other hand, for Category B, only lighter PBHs are allowed for the opposite reason, i.e., $\Omega_{\rm GW}^{\rm LVK-D}\lesssim\Omega_{\rm GW}(f=25\,{\rm Hz})$ requiring $f_{\rm brk}\lesssim \,25\,{\rm Hz}$ at much higher frequencies. Our parameter space scanning does not account for a correction factor due to the possible variation of effective degrees of freedom, which would make plateau amplitude slightly less. Therefore, in this analysis, the LIGO constraints on the string parameter space are slightly stronger than they should be \cite{ng5}. As such, from Fig.\ref{fig:fig4}, it seems that without PBH the NANOGrav 2$\sigma$ region is excluded by LIGO-O3. But when the correction factor is considered, a small portion of $2\sigma$ region around $G\mu\sim 10^{-7}$ does not contradict the LIGO-O3 bound \cite{ng5}. Therefore, one can produce Category-B signals for metastable strings with  $M_{\rm BH}\sim 10^{5.5}\,g$. 

\subsection{Numerical results: Expected Signal-to-Noise-Ratio at the Einstein Telescope} 
{\color{black}The standard way to assess the detectability of predicted GW signal in the future planned detectors is to compute the Signal-to-Noise-Ratio (SNR)~\cite{Maggiore:1999vm,Schmitz:2020syl,Saikawa:2018rcs,Allen:1997ad,Romano:2016dpx,Kudoh:2005as}.  Because the PTA GWs are of strong amplitudes, a generic expectation from any such signal spanning a wide frequency range would be to obtain a large SNR at the high-frequency detectors. In the present case, because PBH domination reduces the signal amplitude at higher frequencies, it is useful to quantify the detectability of such signals with SNR. Note that, in our case, the higher the frequency, the smaller the amplitude of GWs. One gets the most reduced signal at higher frequencies for Category C. We assess those signals by computing SNR at the Einstein Telescope (ET). We quantify the SNR using the most general (for arbitrary signal strength) optimal filter function \cite{Kudoh:2005as}
\bea
\tilde{Q}(f)=\frac{U(f)V(f)^*-U(f)^*W(f)}{|V(f)|^2-W(f)^2}
\eea
that returns SNR as \cite{Kudoh:2005as}
\bea
{\rm SNR}=\sqrt{2T_{\rm obs}}\left[\int_{-\infty}^{+\infty}\frac{df}{2}\frac{|U(f)|^2}{|U(f)|^2+W(f)}\right]^{1/2},\label{snrgen}
\eea
where the frequency-dependent functions are given by 
\bea
U(f) &=& C_{IJ}(f) + \delta_{IJ} N_I(f),\\
V(f) &=& C_{IJ}^2 + \delta_{IJ} N_I \big(N_I + 2C_{II}\big),\\
W(f) &=& C_{II}C_{JJ} + C_{II}N_J + C_{JJ}N_I + N_I N_J
\eea
with 
\bea
C_{IJ}(f; t, t) &=& \int \frac{d\Omega}{4\pi} S_h(f) F_{IJ}(f, \Omega; t, t),\\
\braket{\tilde{n}^*_I(f) \tilde{n}_J(f')} &=& \frac{1}{2} \delta_{IJ} \delta(f - f') N_I(f)\\
\langle \tilde{h}^*_A(f, \Omega) \tilde{h}_{A'}(f', \Omega') \rangle &=&
\frac{1}{2} \delta(f - f') \delta^2(\Omega, \Omega') \frac{\delta_{AA'}}{4\pi} S_h(|f|, \Omega).
\eea

The indices $ I, J $ represent the different detectors used to detect the signal, and $ F_{IJ} $ is the antenna pattern function, which characterizes the overlap in the detectors' responses to the signal (related to the overlap reduction function \cite{Romano:2016dpx}). The term $ \tilde{n} $ denotes the noise, while $ \tilde{h} $ represents the GW signal. The indices $ A, A' $ indicate the polarization states of the GW signal, and $ \Omega $ denotes the direction from which the GW originates. The strain power spectral density $ S_h(f) $ is related to the $ \Omega_{\text{GW}}(f) $ as
\bea
\Omega_{\text{GW}}(f) = \frac{4\pi}{3} \frac{f^3 S_h(f)}{H_0^2}.\label{gwstrn}
\eea
Defining $ C_{IJ}\equiv  S_h(f)\Gamma_{IJ}$, for ET detector network SNR in Eq.\eqref{snrgen} simplifies to 

\bea
{\rm SNR}_{\rm ET} =\sqrt{T_{\rm obs}}\left[ \int_{f_{\rm min}}^{f_{\rm max}}df \left(\frac{\Omega_{\rm GW}(f)^2}{\frac{2}{3}\Omega_{\rm GW}(f)^2+\frac{2}{\sqrt{3}}\Omega_{\rm GW}(f)\Omega_{\rm noise}(f)+\Omega_{\rm noise}(f)^2}\right)\right]^{1/2}\,,\label{snr}
\eea
where we have assumed equal noise for all the detectors and defined $S_n(f)=\frac{N}{\sqrt{3}\Gamma_{\rm ET}}$ \cite{Schmitz:2020syl}, which is related to $\Omega_{\rm noise}$ following Eq.\eqref{gwstrn}.  Note that the SNR defined in Eq.\eqref{snr}, reduces to the standard formula \cite{Maggiore:1999vm,Schmitz:2020syl} used vastly in the literature in the weak signal limit ($\Omega_{\rm GW}\ll \Omega_{\rm noise}$). In Fig.\ref{fig:fig5}, we present the ${\rm SNR}_{\rm ET}$ for PTA GWs radiated from superstrings (metastable strings) in the upper (lower) panel for the benchmark values $M_{\rm BH}=10^{7,8}$ g. Note that even though at the ET band the signal strength reduces, the predicted SNR is pretty high. This is owing to the required strong amplitude GWs at the PTA band plus the maximum value of PBH mass, around $M_{\rm BH}\sim 10^{8}\, g-$the BBN constraint. 

\begin{figure}
    \centering
    \includegraphics[scale=0.65]{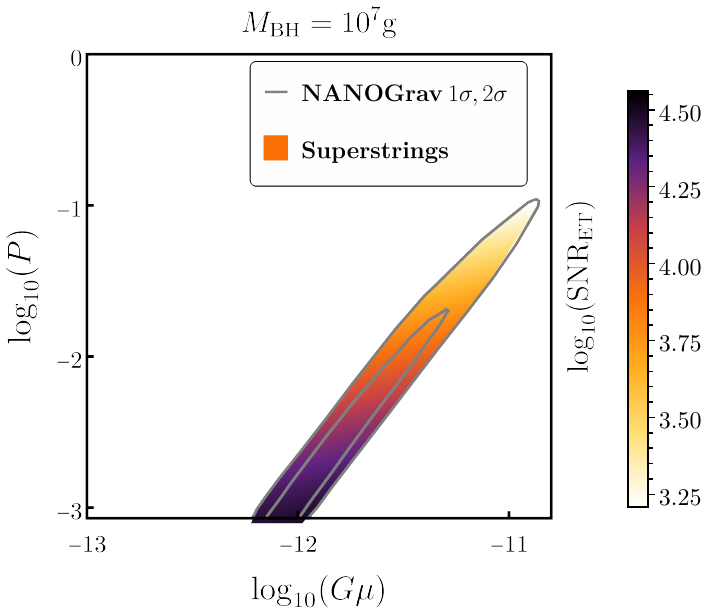}
    \includegraphics[scale=0.65]{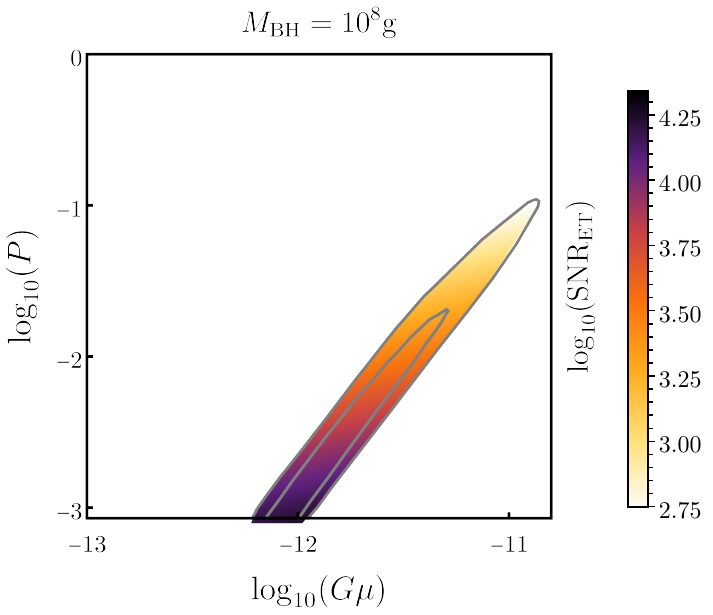}\\
     \includegraphics[scale=0.65]{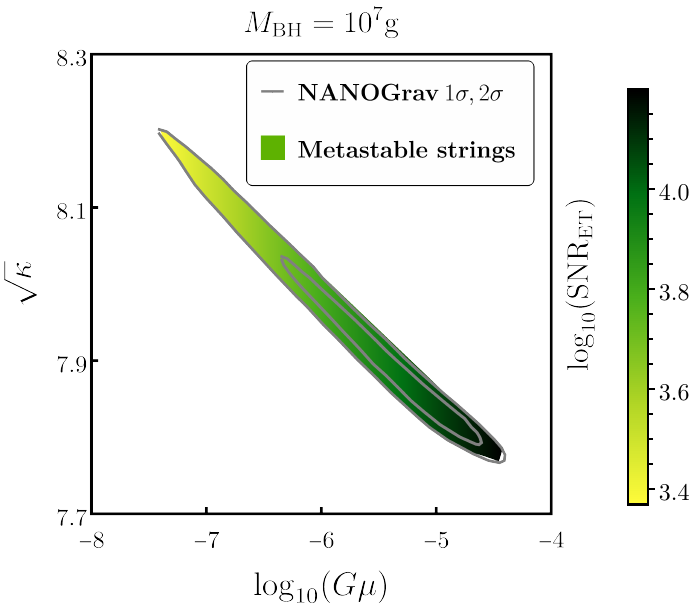}
    \includegraphics[scale=0.65]{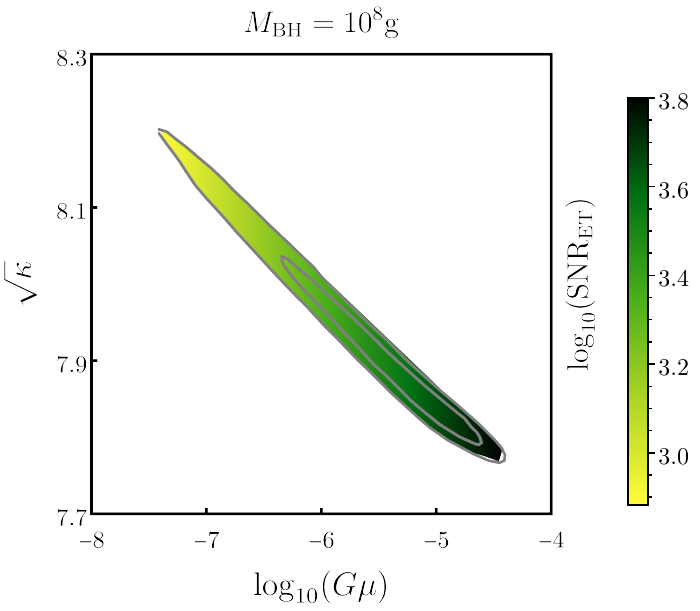}
    \caption{Top [bottom]: Expected Signal-to-Noise-Ratio (SNR) at the Einstein Telescope for the PTA GWs radiated from cosmic superstrings [metastable strings] for $M_{\rm BH}\sim 10^{7}\, g$ (left), and $M_{\rm BH}\sim 10^{8}\, g$ (right). The SNR in each plot has been computed considering a $f^{-1/3}$ fall of the GWs beyond $f_{\rm brk}$. }
    \label{fig:fig5}
\end{figure}
Given the above discussion and the parameter scan rendering an approximate (fairly accurate) range of PBH masses that make string-radiated GWs consistent with the LIGO bound, we now discuss the phenomenological advantage of having a PBH domination over any other early matter domination. }

\section{ PBHs over any other EMD: additional physics }\label{s4}
As mentioned in the introduction, the advantage is broadly twofold. First, PBHs themselves are additional sources of GWs, thereby making the overall stochastic GW spectrum peculiar. As such, PBHs inevitably produce high-frequency gravitons, plus, as shown recently, the inhomogeneity in the distribution of PBHs over the spaces could induce tensor modes if the PBHs dominate the universe's energy density. As we show below, remarkably, for the obtained PBH mass range in Fig.\ref{fig:fig3} and Fig.\ref{fig:fig4}, the peak of such induced GWs from  PBH density fluctuations appears in the LISA-DECIGO-ET band. Second, lately, there have been considerable efforts to discuss the effect of ultralight PBHs on particle dark matter and baryogenesis models \cite{Fujita:2014hha,Lennon:2017tqq,Hooper:2019gtx,Morrison:2018xla,Perez-Gonzalez:2020vnz,Bernal:2022pue,Cheek:2021cfe,Gondolo:2020uqv,Morrison:2018xla,Franciolini:2023osw,Barman:2022gjo,Barman:2022pdo,Borah:2022vsu,Barman:2024slw,Borah:2024qyo,Calabrese:2023bxz,Schmitz:2023pfy,Choi:2023kxo,Calabrese:2023key,Gehrman:2023esa,Bhaumik:2022pil,Bernal:2021bbv,Wu:2021gtd,Sandick:2021gew,Bernal:2021yyb,Cheek:2021odj,Masina:2021zpu,Ambrosone:2021lsx,Kitabayashi:2021hox,Khlopov:2004tn}. The PBHs with the predicted mass range in this article would have profound effects (perhaps the best testable effects)  on BSM particle physics models. To this end, we discuss a recently proposed BSM scenario \cite{Borah:2022iym} based on the seesaw mechanism \cite{Gell-Mann:1979vob} of neutrino masses stemming also the possibility of baryogenesis via leptogenesis \cite{Fukugita:1986hr} and superheavy dark matter in the presence of PBHs \cite{Samanta:2021mdm,Borah:2022iym}. In the scenario, it was reported for the first time, that if there are ultralight PBHs in the early universe, seesaw Lagrangian can accommodate the novel right-handed neutrino (RHN) mass spectrum $\Lambda_{\rm GUT}\sim M_{R3}>M_{R2}>M_{R1}$, where $M_{Ri}$ being the mass of the $i$th RHN, with heaviest one ($N_{R3}$) being super heavy dark matter and the other two generating baryon asymmetry. The analysis in Ref.\cite{Borah:2022iym}  was made with stable cosmic strings produced owing to the spontaneous breaking of an $U(1)_{B-L}$ (embeddable in GUT) responsible also for generating the RHN masses. Here we present an analysis of the same model considering a metastable string network incorporating the new PTA data. In what follows, we present a concise discussion with numerical results for the additional GWs from PBHs and the $U(1)_{B-L}^{\rm PBH}$ seesaw model. 
\subsection{Complementary GW signatures: density fluctuations and graviton emission}\label{s4a}
\begin{figure}
    \centering
    \includegraphics[scale=0.75]{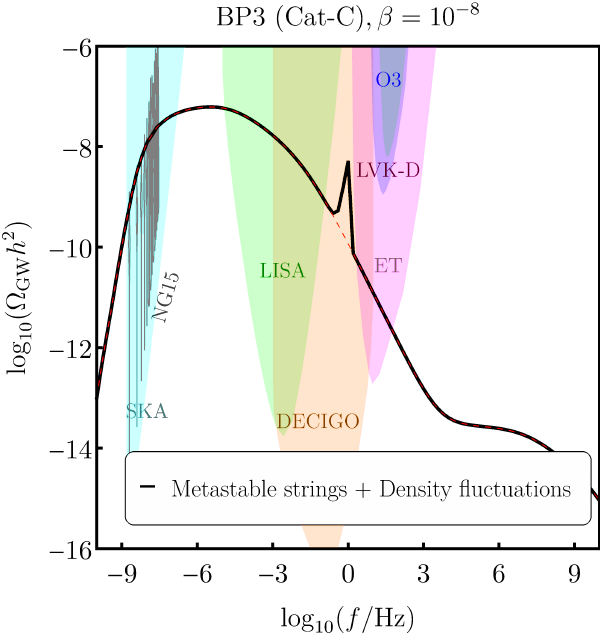}\includegraphics[scale=0.75]{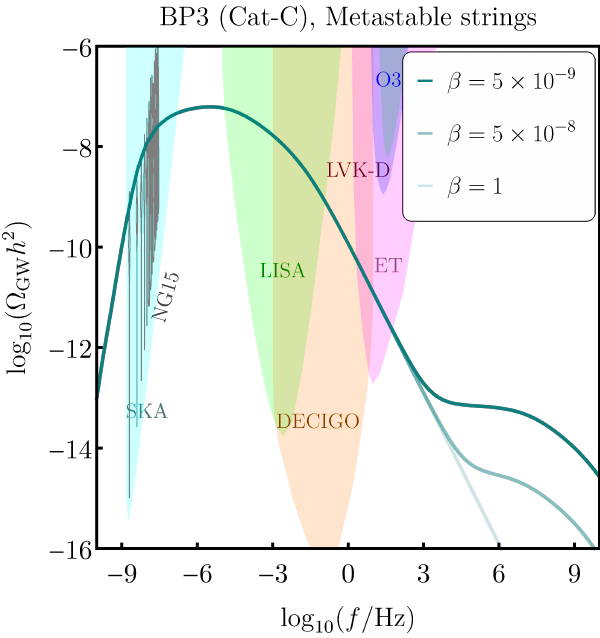}\\
     \includegraphics[scale=1.2]{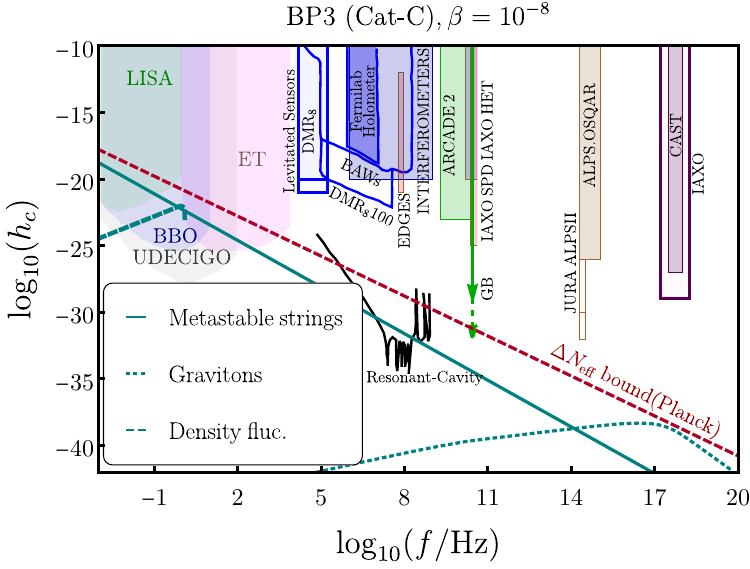}
    \caption{Top-left: Combined GW spectrum (Metastable strings + density fluctuations) for the benchmark parameters belonging to Category C, see Table \ref{t2}. The spectrum has been produced for $\beta= 10^{-8}$. Top-right: GWs only from metastable cosmic strings for the same benchmark parameters but for different values of $\beta$, showing the impact of the finite duration of PBH domination. Bottom: GW strain $h_c$ from density fluctuations (dashed), from Metastable strings (solid), and graviton radiation from PBH (dotted) with the projected sensitivities  (including the black curly curve) of various high-frequency GW detectors. All the benchmark parameters are the same as in the top-left plot. The region above the red dashed line is excluded from $\Delta N_{\rm eff}$ bound by Planck.  }
    \label{fig:fig6}
\end{figure}
\begin{figure}
    \centering
    \includegraphics[scale=0.75]{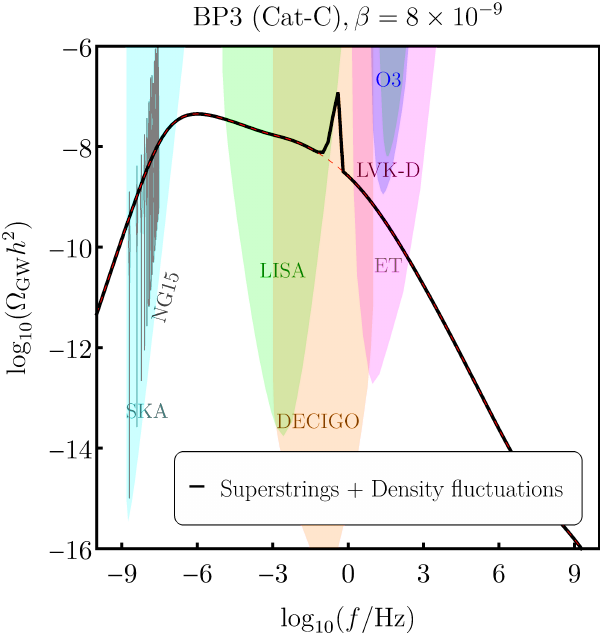}\includegraphics[scale=0.75]{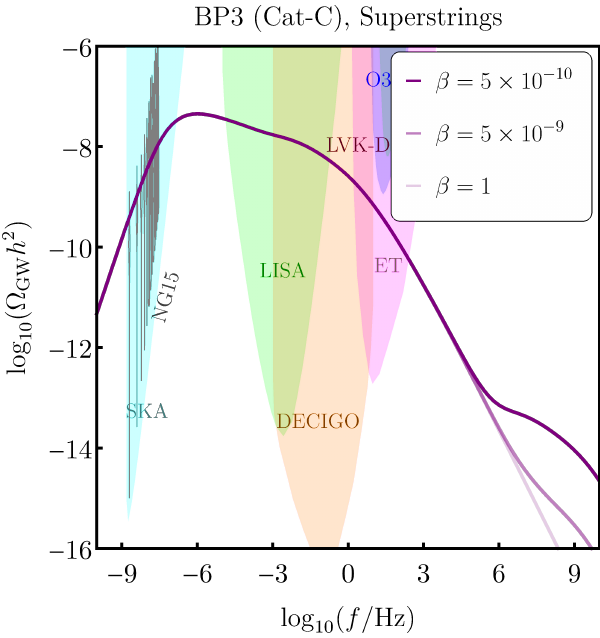}\\
     \includegraphics[scale=1.2]{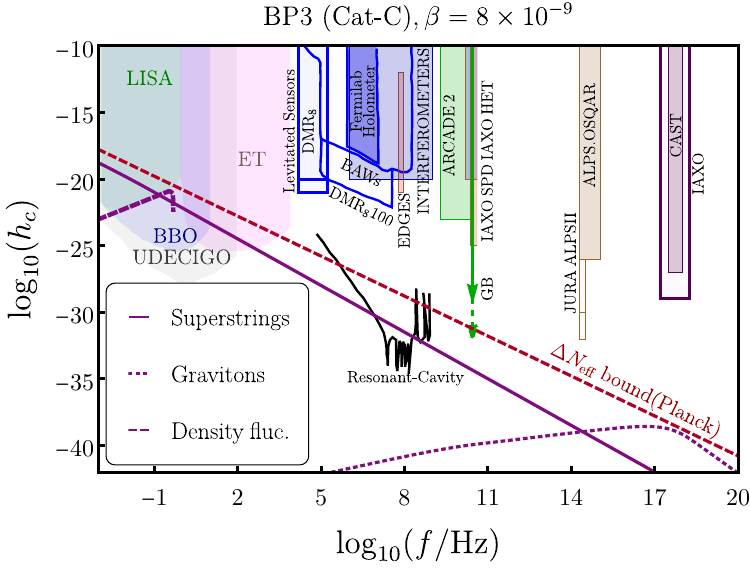}
    \caption{Top-left: Combined GW spectrum (Superstrings + density fluctuations) for the benchmark parameters belonging to Category C, see Table \ref{t1}. The spectrum has been produced for $\beta=8\times 10^{-9}$. Top-right: GWs only from superstrings for the same benchmark parameters but for different values of $\beta$, showing the impact of the finite duration of PBH domination. Bottom: GW strain $h_c$ from density fluctuations (dashed), from superstrings (solid), and graviton radiation from PBH (dotted) with the projected sensitivities  (including the black curly curve) of various high-frequency GW detectors. All the benchmark parameters are the same as in the top-left plot. The region above the red dashed line is excluded from $\Delta N_{\rm eff}$ bound by Planck.}
    \label{fig:fig7}
\end{figure}

The ultralight PBHs are many-faced objects concerning the stochastic GWs. For instance, the initial curvature perturbations leading to the formation of PBHs give rise to GWs (see, e.g., \cite{pbhf1,pbhf2,pbhf3,Martin:2019nuw}), ultralight PBHs could possibly merge, releasing  GWs \cite{Hooper:2020evu}, PBHs emit gravitons, which constitute GWs at ultra-high frequencies \cite{Anantua:2008am,Dolgov:2011cq,Dong:2015yjs,Ireland:2023avg}, and the inhomogeneous distribution of PBHs, which results in density fluctuations, triggers the production of GWs \cite{Papanikolaou:2020qtd,Domenech:2020ssp,Domenech:2021wkk}. Here we do not discuss any specific production mechanisms of PBHs, nor the possibility of merger formation. We rather focus on the last two options, being slightly biased to the fact that graviton emission is inevitable and independent of the production mechanism, the GWs from PBH-density fluctuations seem to provide a realistic stochastic background with spectral shape dependent on their spatial distribution which has been assumed to be like Poisson like in Refs.\cite{Papanikolaou:2020qtd,Domenech:2020ssp}. Even though the PBH gas, on average, behaves like pressure-less dust, the inhomogeneous spatial distribution results in density fluctuations, which are isocurvature in nature. When PBHs dominate the universe's energy density, the isocurvature component gets converted to curvature perturbation, which subsequently induces secondary GWs. Due to the significant density fluctuations at smaller scales (equivalent to the average separation of PBHs at $T_{\rm Bf}$), substantial GWs are generated. These GWs are further amplified due to the nearly instantaneous evaporation of PBHs. The amplitude of such induced GWs in the present day is given by \cite{Domenech:2021ztg}
\begin{equation}
\Omega_{\rm GW}^{\rm PBH,I}(t_0,f)\simeq \Omega^{\rm peak,I}_{\rm GW} \left( \frac{f}{f_{\rm peak}}\right)^{11/3} \Theta (f_{\rm peak,I}-f),
\end{equation}
where the peak amplitude is given by
\begin{equation}
\Omega^{\rm peak,I}_{\rm GW}\simeq 2\times 10^{-6}\left(\frac{\beta}{10^{-8}}\right)^{16/3} \left(\frac{M_{\rm BH}}{10^7 g}\right)^{34/9},\label{peakamp}
\end{equation}
and the peak frequency is given by
\begin{equation}
f_{\rm peak,I}\simeq 5.4 \text{\:Hz}\left( \frac{M_{\rm BH}}{10^7 g}\right)^{-5/6}.
\end{equation}
The ultra-violet cutoff frequency $f_{\rm peak,I}$ corresponds to the comoving scale representing the mean separation of PBHs at the formation time.

On the other hand, the present spectral energy density of GWs from gravitons is given by \cite{Anantua:2008am,Dolgov:2011cq,Dong:2015yjs}
\begin{equation}
\Omega_{\rm GW}(f) h^2= \frac{(2\pi)^4\beta \gamma^2f^4}{4H_0^2M_{\rm BH}}\int_{t_{\rm Bf}}^{t_{\rm ev}} \frac{(1-\frac{t-t_{\rm Bf}}{\tau_{\rm ev}})^{2/3}\left( \frac{a(t_{\rm Bf})}{a(t_0)}\right)^{-3}}{\exp\left( \frac{16\pi^2 f a(t_0)M_{\rm BH}}{a(t)M_{\rm Pl}^2}\right)-1}\label{graviton}
\end{equation}
By maximizing the Eq.\eqref{graviton} w.r.t the present-day frequency $f$, one obtains the peak frequency and consequently the peak GW amplitude as 

\bea
f_{\rm peak,g}\simeq 2.8 a(t_{\rm ev})T_{\rm BH}\simeq  2 \times 10^{17}\text{\:Hz}\left( \frac{M_{\rm BH}}{10^7 g}\right)^{1/2},\,\,\Omega_{\rm GW}^{\rm peak,g}\sim 10^{-7}.\label{graviton1}
\eea
In Fig.\ref{fig:fig6} and Fig.\ref{fig:fig7} (top-left) we show the total GW spectrum: density fluctuations spectrum combined with the one from metastable strings and superstrings for the benchmark parameters belonging to Category C in each string class. Instead of large $\beta$, here we chose a benchmark value such that the duration of PBH domination is relatively short and the density fluctuations GW spectrum is visible (the sharp peak at higher frequencies). 
Note that because of the strong dependence of the peak GW amplitude on $\beta$ (see Eq.\eqref{peakamp}) in this scenario, the duration of PBH domination can not be arbitrarily large, otherwise, the GWs would saturate the BBN bound on $\Delta N_{\rm eff}$: $\int{\rm d} f ~f^{-1}\Omega_{\rm GW}(f)h^2\lesssim 5.6\times 10^{-6}\Delta N_{\rm eff}$, with $\Delta N_{\rm eff}\lesssim 0.2$. Consequently, the falling of the GW spectrum from cosmic strings would stop at higher frequencies, potentially showing up a subsequent GW plateau originating from the loops in the radiation epoch before the PBH domination. This has been shown in the top-right panel of Fig.\ref{fig:fig6} for metastable strings for three benchmark values of $\beta$. Nonetheless, the appearance of the second plateau is not always the case if the particle production cut-off is strong enough. We show it for superstrings on the top-right panel of Fig.\ref{fig:fig7}. Compared to the metastable string case, here owing to the strong particle production cut-off, the GW spectra closely follow the large $\beta$  case. As discussed in Sec.\ref{s2}, for the GWs with amplitude at the level of PTAs, the particle production cut-off appears mostly at frequencies higher than the LIGO frequency band. Therefore, for extremely shorter-duration PBH-domination scenarios, the regions in Fig.\ref{fig:fig3} and Fig.\ref{fig:fig4} extracted for $\Omega_{\rm GW}\propto f^{-1/3}$ falling, requires modification now to account for a flattening of the spectrum. Although we do not present such a scan, to observe GW peak from density fluctuations on top of the strong amplitude cosmic string radiated GWs required by PTAs, reasonably large $\beta$ (respecting, of course, the BBN bound) is required. For this reason, for most of the parameter space presented in Fig.\ref{fig:fig3} and Fig.\ref{fig:fig4}, the duration of PBH domination would be such that from $f_{\rm brk}$ until the $f_{\rm LIGO}$, the spectrum would fall as $\Omega_{\rm GW}\propto f^{-1/3}$, keeping the results of the previous section intact. We expect modification to the parameter space for heavier PBH masses around $10^8$ g leading to smaller values of $f_{\rm brk}$, and thereby potential flatting of the spectrum at the LIGO frequency band due to the shorter duration of PBH domination. 

Besides these low-frequency peculiarities, a PBH+superstrings (metastable strings) system provides a characteristic high-frequency GW spectrum constituting gravitons as shown in the bottom panels of Fig.\ref{fig:fig6} and \ref{fig:fig7}, where the GW strain $h_c$ is related to $\Omega_{\rm GW}$ as
\bea
h_c\left(f\right)=\sqrt{\frac{3 H_0^2}{2\pi^2}\Omega_{\rm GW}}f^{-1}.
\eea
Because Eq.\eqref{graviton1} characterizes the GW background, for the allowed mass ranges of PBH presented in Sec.\ref{s3}, a considerable GW background constituting gravitons can be obtained within a constrained frequency range spanning approximately, only a decade. Although the proposed high-frequency GW detectors \cite{hf1,hf2,hf3,hf4,hf5,hf6,hf7,hf8,hf9,hf10,hf11} are not presently competitive with the $\Delta N_{\rm eff}$ bound from Planck \cite{Planck} (according to Ref.\cite{hf12}, the electromagnetic resonant cavities can potentially reach below
the dark radiation projections around $f\sim 10^8$ Hz; the black oscillating curve in Fig.\ref{fig:fig6} and Fig.\ref{fig:fig7}), optimistically, in the future, graviton background may be detected. In any case, these marked spectral features in the total GW background are described with three characteristic frequencies $f_{\rm brk}$, $f_{\rm peak,I}$, and  $f_{\rm peak,g}$, making the PBH + string scenario more appealing than an ordinary EMD + string case. 
\subsection{Implication on a realistic BSM model of leptogenesis and super-heavy dark matter}\label{s4b}
\begin{figure}
    \centering
    \includegraphics[width=1\linewidth]{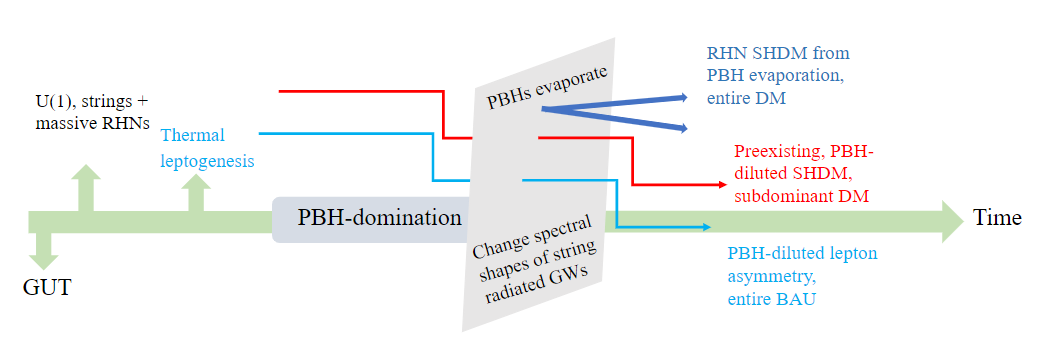}
    \caption{A tentative timeline for the events occurring in the $U(1)_{B-L}^{\rm PBH}$ model. Without discussing any GUT scheme, this article assumes a metastable string network formation after the $U(1)_{B-L}$ breaking which also makes the right-handed neutrinos (one of them being a super-heavy dark matter) massive.}
    \label{fig:timeline}
\end{figure}
In the seesaw mechanism \cite{Gell-Mann:1979vob}, the light neutrino mass $m_i\sim 0.01$ eV inferred by neutrino oscillation data and cosmological measurements \cite{Planck,nufit}, can be explained as a ratio of two scales: $m_i\simeq \frac{ \Lambda_{\rm EW}^2}{\Lambda_{\rm GUT}}$, with $\Lambda_{\rm EW}$ and $\Lambda_{\rm GUT}$ being the Electroweak and GUT scale, respectively. At the Lagrangian level, these two scales are introduced as $\mathcal{L}_{\rm seesaw}\sim  \Lambda_{\rm EW} \overline{L} N_R+\Lambda_{\rm GUT} \overline{N^c_R} N_R$, where $L$ is the Standard Model (SM) lepton doublet and $N_R$ is a RHN field. Besides generating light neutrino mass, the seesaw Lagrangian can explain the entire DM relic and the observed BAU, if there are three RHN fields, with one being the DM and the other two generating BAU through leptogenesis.
A plethora of studies is dedicated to exploring this three-dimensional aspect of the seesaw--the neutrino mass + BAU + DM, see, e.g., \cite{uni1,uni2,uni3,uni4,uni5,uni6,uni7}. Such attempts are theoretically well-built and predictive in particle physics experiments, however, at the expense of fine-tuning the $\overline{L} N_R$ 
 Yukawa coupling and treating the RHN mass as a free parameter, thereby giving up the supposition that active neutrino masses arise from the GUT and EW scale physics. As such, perhaps the most studied scenarios based on \cite{uni1,uni2} consider $N_R$ to be extremely light $\mathcal{O}(\rm GeV,\, keV)$, therefore tiny $\overline{L} N_R$ coupling. In Ref.\cite{Borah:2022iym}, it was proposed that if there are PBHs in the early universe, all three aspects of the seesaw can be addressed without significant deviation from the relation $m_i\simeq \frac{ \Lambda_{\rm EW}^2}{\Lambda_{\rm GUT}}$, i.e., considering all the RHNs to be $\mathcal{O}(\Lambda_{\rm GUT})$, with one of them being the GUT-scale (super heavy) DM. In this case, the scenario becomes testable in GW experiments. The idea is based on dark matter production from ultralight PBH evaporation \cite{Fujita:2014hha,Lennon:2017tqq,Hooper:2019gtx,Morrison:2018xla}. If such DM makes up  100$\%$ relic density without contradicting structure formation, the DM mass must be in the super-heavy range following the relation 
\bea
M_{\rm DM}\simeq 4.7\times10^{12} \,\left(\frac{M_{\rm BH}}{10^7 g}\right)^{-5/2}\,{\rm GeV}.\label{PBHDM}
\eea
The thermal leptogenesis mechanism (thermal scatterings populate the decaying RHNs producing the lepton asymmetry) generates the BAU.  But in this case, the final asymmetry gets modified to account for an entropy dilution owing to the PBH domination. Once the condition for the observed baryon-to-photon ratio $\eta_B\simeq 6.3\times 10^{-10}$ \cite{Planck} is taken into account, the parameter $\beta$ and scale of thermal leptogenesis get correlated as \cite{Borah:2022iym} 
\bea
\beta\simeq 1.2\times 10^{-9}\left(\frac{10^7g}{M_{\rm BH}}\right) \left(\frac{T_{\rm lepto}}{10^{14}\rm GeV}\right),\label{lepto}
\eea
where we have assumed leptogenesis is dominated by the decay of $N_1$ so that $M_1\sim T_{\rm lepto}$. Eq.\eqref{lepto} can be shown to be accurate, matching the numerical results obtained from Eq.\eqref{den1},\eqref{den2} and the modified leptogenesis equations in the presence of PBHs \cite{csfit4}. The PBHs can also populate the decaying RHNs to produce lepton asymmetry, but if Eq.\eqref{PBHDM} is to be satisfied, such PBHs evaporate after Sphaleron freeze-out temperature, and the lepton asymmetry does not get processed to BAU. Therefore we discard this option. 
\begin{figure}
    \centering
    \includegraphics[scale=0.85]{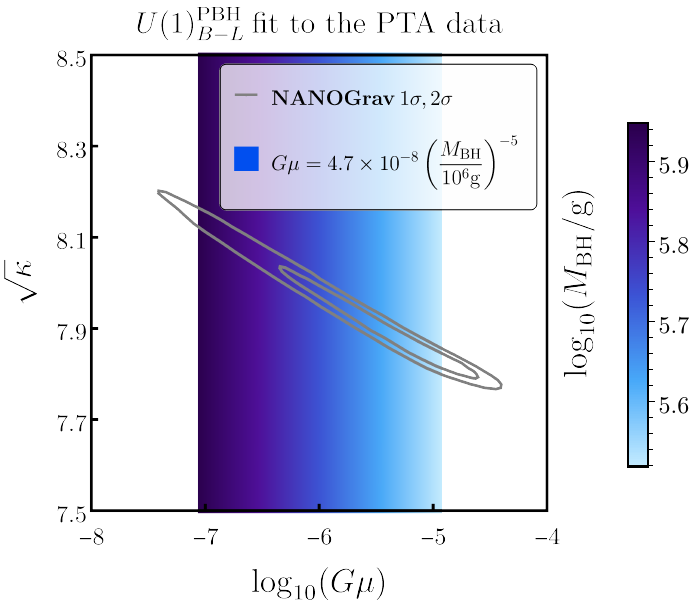}
    \caption{Fit to the PTA data with $U(1)_{B-L}^{\rm PBH}$ model for neutrino mass, super-heavy DM and leptogenesis. The produced GW signal from metastable strings belongs to Category A discussed in the previous section, i.e., the GW spectrum consistent with the PTA data must produce a signal at the LVK-D.  }
    \label{fig:DMfit}
\end{figure}
Because the seesaw Lagrangian can be embedded in GUT groups with the RHN mass term generated after the breaking of an intermediate (While going to ${\rm GUT\rightarrow SM}$)  symmetry $U(1)_{B-L}$, cosmic strings are formed. Not exploring any explicit GUT scheme, we consider the formation of a metastable string network after the  $U(1)_{B-L}$ breaking. The tentative timeline of this $U(1)_{B-L}$  model in the presence of PBHs looks like Fig.\ref{fig:timeline}. It is crucial to note that like the BAU, dilution of any pre-existing DM relic (e.g., from scattering by $B-L$ gauge bosons) should be taken care of (the red line in Fig.\ref{fig:timeline}) in a way that Eq.\eqref{PBHDM} prevails. A detailed discussion related to this can be found in Ref.\cite{Borah:2022iym}. Note now the features of the model. First, The DM and the PBH masses are anti-correlated (cf. Eq.\eqref{PBHDM}). As such, $\sim 10^{15}$ GeV DM gets produced from $\sim 10^6\,g$ PBHs. Lighter PBHs are required to produce heavier DM.  Second, as the leptogenesis scale increases (cf. Eq.\eqref{lepto}), $\beta$ increases, consequently the duration of PBH domination. Therefore, as the RHN mass scale (including the DM) approaches $\Lambda_{\rm GUT}$,  remarkably, besides maintaining the  $m_i\simeq \frac{ \Lambda_{\rm EW}^2}{\Lambda_{\rm GUT}}$ relation to a great extent, the GWs produced from cosmic strings automatically evades LIGO bound owing to large $\beta$ or the long duration of PBH-domination. Assuming the hierarchy in the RHN masses\footnote{This assumption is not necessary, but more convenient from the model building perspective. As such, one can choose to impose a simple condition  $M_{\rm DM}>T_{\rm RH}>M_{Ri(=1,2)}$ on the parameter space, to avoid DM production from scattering while maintaining the condition for thermal leptogenesis $T_{\rm RH}>M_{Ri(=1,2)}$. } as $M_{\rm DM}>M_{Ri(=1,2)}$ we can re-write Eq.\eqref{tension} as
\bea
\mu=\pi M_{\rm DM}^2~f_{\rm DM}^{-2}h\left(\lambda,g^\prime\right),\label{tensionDM}
\eea
where $f_{\rm DM}$ is the DM Yukawa coupling which determines the interaction strength of $N_R$ field with $B-L$ scalar; $\overline{N^c_R} N_R\Phi_{B-L}$. Working with $\mathcal{O}(1)$ couplings in Eq.\eqref{tensionDM}, and using Eq.\eqref{PBHDM} we obtain an expression for $G\mu$ as 
\bea
G\mu\simeq 4.7\times 10^{-8}\left(\frac{M_{\rm BH}}{10^6 g}\right)^{-5}. \label{cond}
\eea
With this relation, we now redo the PTA fit for metastable strings performed in the previous section. The result is shown in Fig.\ref{fig:DMfit}. Note that because of this extra condition in Eq.\eqref{cond}, this $U(1)_{B-L}^{\rm PBH}$ model fits the PTA data for an extremely constrained range of PBH masses and produces  GW signals belonging only to Category A. Therefore any non-observation of SGWB in the LVK-D would rule out the model/the best parameter choice of the model. Besides, because the PBH mass window is narrow, the other GW signals, e.g., from density fluctuations plus in the form of gravitons are peaked in a stringent range in frequencies. In an upcoming publication, we shall provide a detailed discussion of the model for more general parameter space with all possible GW signatures.

\section{Conclusion}
An abundant presence of ultralight PBHs leading to their domination on energy densities in the early universe could produce plenty of observable signatures specifically in the form of gravitational radiation. In addition to producing direct GW background, e.g., in the form of gravitons, a PBH domination gets imprinted on the spectral shape of GWs originating from other independent sources. Such a spectral shape reconstruction with the GW detectors and studying complementary GW signals originating directly from the PBHs could be an interesting way to probe the properties of ultralight PBHs. Here we study the impact of PBH domination on the GWs emitted from cosmic superstrings and metastable strings which are speculated to be among the sources of nHz GWs observed by the PTAs. Such an interpretation might become perilous, as a GW background originating from cosmic strings spans a wide range in frequencies thereby contradicting the non-observation of stochastic GW background by LIGO dealing with comparable signal strength. In fact, the LIGO constraint rules out a significant string-parameter space favoured by the recent PTA data. A PBH-dominated early universe, however, affects the propagation of such GWs to reduce the amplitude at the high frequencies thereby evading the LIGO constraints. Assuming monochromatic and non-rotating PBHs we work out the mass range of such ultralight PBHs which makes the string-radiated GWs consistent with LIGO bound and the PTA data. We show that PBHs with mass in the range $M_{\rm BH}\in \rm \left[10^6g,10^8 g\right]$ would produce such viable signals that can be tested in the mid-band detectors such as LISA, high-frequency detectors such as LVK-D plus ET with a high signal-to-noise ratio. We discuss why a PBH-domination is distinguishable from any other early matter-domination scenarios that reduce the amplitude of string-radiated GWs at higher frequencies. In this context, we discuss a BSM model based on $U(1)_{B-L}$-seesaw providing neutrino mass, leptogenesis, super-heavy dark matter, and metastable cosmic strings. We show how a PBH-cosmic string scenario could be extremely predictive in probing unexplored parameter space of such BSM scenarios. 
\section*{Acknowledgements}
 The work of SD is supported by the National Natural Science Foundation of China (NNSFC) under grant No. 12150610460. The work of  RS is supported by the research project TAsP (Theoretical Astroparticle Physics) funded by the Istituto Nazionale di Fisica Nucleare (INFN).
 \appendix
 \section{Superstrings and Metastable strings fit to the NANOGrav data}\label{a1}
\begin{figure}
    \centering
    \includegraphics[scale=0.8]{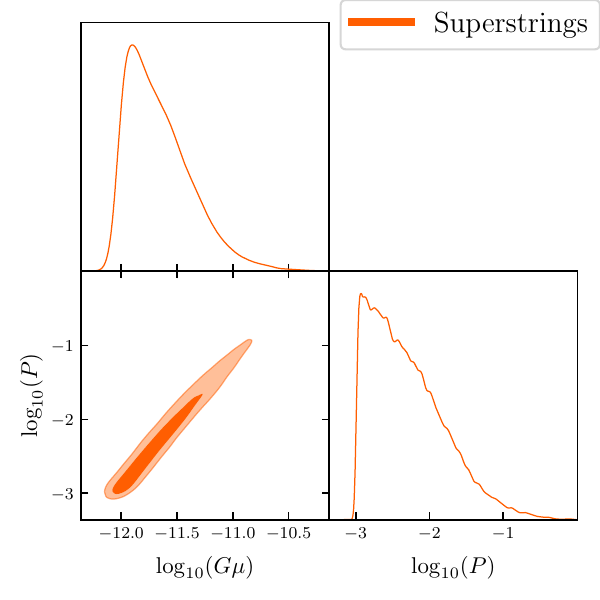}\includegraphics[scale=0.8]{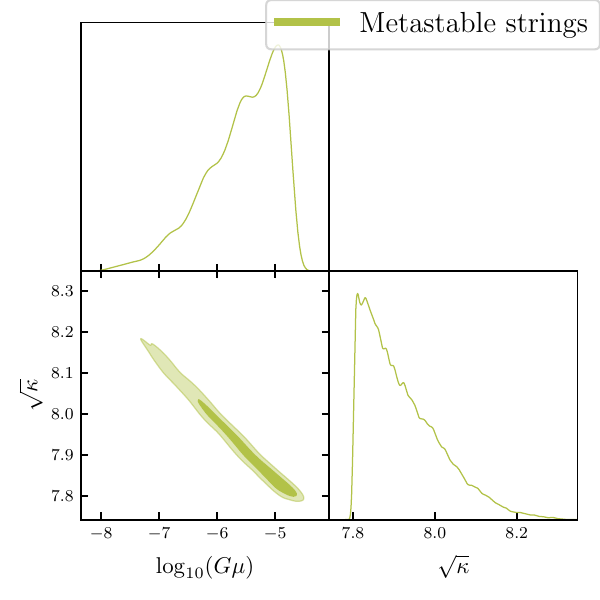}\\
    \caption{ Triangular plot on the left (right) from the MCMC runs for superstrings (metastable strings) with the NANOGrav 15-year dataset using \texttt{PTArcade}\cite{Mitridate:2023oar} in the \texttt{ceffyl} mode\cite{Lamb:2023jls}.}
    \label{fig:fig10}
\end{figure}

 In Fig.\ref{fig:fig10} we present the superstring and metastable fit to the PTA data using $\texttt{PTArcade}$ \cite{Mitridate:2023oar}. The 2D posteriors have been used in the main text.
 \bibliography{bibliography}
\end{document}